\documentclass[runningheads]{llncs}
\newif\ifdraft
\draftfalse

\newif\ifanonymous
\anonymousfalse

\input{macros-generic}

\usepackage{graphicx}
\usepackage{tikz}
\usepackage{booktabs}
\usetikzlibrary{arrows.meta,positioning,fit}
\usepackage{xspace}
\usepackage[acronym,toc]{glossaries}
\usepackage[capitalize,noabbrev]{cleveref}
\usepackage{adjustbox}
\usepackage[normalem]{ulem}
\usepackage{paralist}
\usepackage{multirow}
\usepackage{url} 

\renewcommand{\paragraph}[1]{ \noindent \textbf{#1.~}\xspace}

\begin{document}
\title{Cross-Layer Deanonymization Methods in the Lightning Protocol}

%
%
\ifanonymous
\author{Anonymous submission\inst{1}}
\institute{Anonymous \\
	\email{anonymous}\\
}
\authorrunning{F. Author et al.}
\else
\author{Matteo Romiti\inst{1} \and
Friedhelm Victor\inst{2} \and
Pedro Moreno-Sanchez\inst{3} \and Peter Sebastian Nordholt\inst{5} \and Bernhard Haslhofer\inst{1}
\and Matteo Maffei\inst{4}}
\authorrunning{M. Romiti et al.}
%
\institute{Austrian Institute of Technology
\email{\{matteo.romiti, bernhard.haslhofer\}@ait.ac.at}\\
\and
Technische Universit\"at Berlin
\email{friedhelm.victor@tu-berlin.de}
\and
IMDEA Software Institute
\email{pedro.moreno@imdea.org}
\and
Technische Universit\"at Wien
\email{matteo.maffei@tuwien.ac.at}
\and
Chainalysis
\email{psn@chainalysis.com}
}
\fi
\maketitle              
\newif\ifshortabs
\shortabstrue
\ifshortabs
\begin{abstract}
  Bitcoin (BTC) pseudonyms (layer 1) can effectively be de\-anon\-ymized using
  heuristic clustering techniques. However, while performing transactions off-chain (layer 2) in the Lightning Network (LN) 
  seems to enhance privacy, a systematic analysis of the anonymity and privacy
  leakages due to the interaction between the two layers is missing.
  We present clustering heuristics that group \Bitcoin addresses, based on their
  interaction with the \LN, as well as \LN nodes, based on
  shared naming and hosting information. We also present linking
  heuristics that link
  \percNodesLinkedHeurTwoMinConfAllAliasIP\% of all \LN nodes to
  \percAddressesLinkedHeurTwoMinConfAll\% \Bitcoin addresses interacting with the
  LN.
  These links allow us to attribute information (e.g., aliases, IP
  addresses) to \percAddressesAliasIPTagged\% of the \Bitcoin addresses
  contributing to their deanonymization. Further, these deanonymization results
  suggest that the security and privacy of \LN payments are weaker than
  commonly believed, with \LN users being at the mercy of as few as five actors
  that control 36 nodes and over $33\%$ of the total capacity. Overall, this is
  the first paper to present a method for linking \LN nodes with \Bitcoin
  addresses across layers and to discuss privacy and security implications.
\end{abstract}
\else
\begin{abstract}
  Bitcoin was initially believed to enable anonymous Layer-1 (i.e., on-chain)
  payments, a claim soon disputed by numerous works introducing effective
  techniques to cluster addresses belonging to the same user. In the last years,
  Layer-2 (i.e., off-chain) solutions, such as payment channel networks (PCNs)
  have gained popularity as a mean to mitigate the scalability issues inherent
  to cryptocurrencies like Bitcoin (BTC). While performing transactions off-chain
  seems to enhance privacy, a systematic analysis of the anonymity and privacy
  leakages due to the interaction between the two layers is missing.
  In this paper, we focus on the \Bitcoin Lightning Network (LN), which is the
  most widespread implementation of PCNs to date. We present clustering
  heuristics that group \Bitcoin addresses, based on their interaction with the
  \LN, as well as \LN nodes, based on shared naming and hosting information. We
  also present cross-layer linking heuristics that can, with our dataset, link
  \percNodesLinkedHeurTwoMinConfAllAliasIP\% of all \LN nodes to
  \percAddressesLinkedHeurTwoMinConfAll\% \Bitcoin addresses interacting with the
  LN.
  These cross-layer links allow us to attribute information (e.g., aliases, IP
  addresses) to \percAddressesAliasIPTagged\% of the \Bitcoin addresses
  contributing to their deanonymization. Further, these deanonymization results
  suggest that the security and privacy of off-chain payments is weaker than
  commonly believed, with \LN users being at the mercy of as few as five actors
  that control 36 nodes and over $33\%$ of the total capacity. Overall, this is
  the first paper to present a method for linking \LN nodes with \Bitcoin
  addresses and to discuss privacy and security implications as well as
  mitigation strategies.
\end{abstract}
\fi
%
%
%


\section{Introduction\label{sec:intro}}

	Payment channel-networks (PCNs) have emerged as a promising alternative to mitigate the scalability issues with current cryptocurrencies. These \lMinusTwo protocols, built on-top of \lMinusOne blockchains, allow users to perform transactions without storing them on the Bitcoin (BTC) blockchain. The idea is that two users create a \openchtx that locks coins, thereby creating a payment channel between them~\cite{GudgeonMRMG19}. Further payments no longer require on-chain \txs but rather peer-to-peer mutual agreements on how to distribute the coins locked in the channel. At any point, both users can decide to close the channel by creating a \closechtx that unlocks the coins and distributes them according to the last agreed balance.

	While there are different payment channel designs, the \Bitcoin Lightning Network (\LN)~\cite{poon_bitcoin_2016} is the most widespread PCN implementation to date. At the time of writing (September 2020), according to 1ml.com, the \LN features a network of \nNodesNow public active \nodes, \nChannelsNow channels and a total capacity of more than \nBTCNow BTC, worth \nUSDNow USD.

	Apart from scalability, PCNs are considered beneficial to improve the well-known lack of privacy of cryptocurrencies~\cite{Decker-privacy}, where the anonymity claim stemming from the usage of pseudonyms in on-chain \txs has been largely refuted from both academia and industry~\cite{Khalilov:2018aa}. The key to an effective deanonymization of \Bitcoin pseudonyms lies in heuristic methods, which cluster addresses that are likely controlled by the same entity~\cite{Meiklejohn2013}. In practice, entities correspond to user wallets or software services (e.g., hosted wallet, exchange) that control private keys on behalf of their users.

	In this work, we challenge the widespread belief that the \LN greatly improves privacy by showing for the first time how  \LN nodes can be linked to \Bitcoin addresses, which results in a bi-directional privacy leakage affecting \LN and \Bitcoin itself. Related research~\cite{Rohrer:2019aa,Seres:2020a,Martinazzi:2020a,Lin:2020a,mariusz_nowostawski_evaluating_2019} already focused on security and privacy aspects on the PCN layer, but, so far, none of them focused on linking off-chain \LN nodes to on-chain \Bitcoin addresses.
	This is a challenging task because such links are not provided explicitly in the \LN protocol as they would severely affect the privacy of node operators (e.g., revealing their business to competitors).

	\paragraph{Our Contributions} Our methodology is structured in two main strategies: (i) heuristics on \lone, to create clusters of \Bitcoin addresses controlled by the same actor, and on \ltwo, clusters of \LN nodes; and (ii) heuristics to link these clusters across layers.
		In \cref{sec:clustering}, we present four novel on-chain clustering heuristics (star, snake, collector, proxy), which group \Bitcoin addresses based on their interaction patterns with the \LN. With these heuristics, we can cluster \percEntitiesFundingClustered\% of all \Bitcoin entities funding an \LN channel, and \percEntitiesSettlementClustered\% of all entities closing a channel. We also present an \LN node clustering heuristic  leveraging public announcements of aliases and IP addresses, which allows us to group \nNodesAliasIPCluster nodes into \nClustersOffChainOnly clusters.
		In~\cref{sec:linking}, we present two novel cross-layer linking algorithms. One exploits that the same \Bitcoin address can be used to close one channel and then re-use the coins to open a new channel, which allows us to link \percNodesLinkedHeurOneAllAliasIP\% of the \LN nodes to \percAddressesLinkedHeurOneAll\% \Bitcoin addresses in our dataset, when combined with the previous on- and off-chain clustering heuristics. The other algorithm exploits the reuse of a single \Bitcoin entity for opening several channels to different \LN nodes and it allows us to link \percAddressesLinkedHeurTwoMinConfAll\% of the addresses  to \percNodesLinkedHeurTwoMinConfAllAliasIP\% of nodes.

		Given these results, we finally discuss the impact of our deanonymization techniques on the privacy of \Bitcoin entities as well as the security and privacy of the \LN. In a nutshell,  we are able to (i) attribute \percAddressesAliasIPTagged\% of the \Bitcoin addresses with information from the \LN (e.g., IP addresses); (ii) measure the centralized control of the capacity in the \LN and observe that as few as five actors consisting of 36 nodes control over $33\%$ of the total capacity; (iii) show that as few as five users can threaten the security of the \LN by means of (possibly targeted) DoS attacks and violate the privacy of over $60\%$ of the cheapest payment paths because they are routed through them.

		For the reproducibility of the results, we make our dataset and our implementation openly available at \url{https://github.com/MatteoRomiti/lightning_study}\footnote{The proprietary attribution data from Chainalysis is not included in the published dataset. The reader can contact the company for further inquiry.}.


\section{Background and Problem Statement}\label{sec:background}

	We now define the simplified model and terminology used throughout this paper, elaborating then on the cross-layer linkage problem, as well as on related work in this area. For further details on PCNs, we refer to recent surveys~\cite{GudgeonMRMG19,Jourenko2019}.

	\subsection{\Bitcoin Blockchain (Layer 1)}\label{sec:background_blockchain}

		A \Bitcoin \textbf{\address} $\addressid$ is a tuple containing (i) a number of coins (in Satoshis) associated to this address; and (ii) an excerpt of the \Bitcoin script language that denotes the (cryptographic) conditions under which $\addressid$ can be used in a transaction. Although in principle it is possible that $\addressid$ can be spent under any condition that can be expressed in the \Bitcoin script language, in practice most of the \addresses share a few conditions: (i) requiring a signature $\sigma$ on the transaction verifiable under a given public key $\pk$; and (ii) requiring two signatures $\set{\sigma_1, \sigma_2}$ verifiable with two given public keys $\pk_1$ and $\pk_2$ (i.e., multisig address). We say that an address $\addressid$ is owned by a \usertext if she can produce the required signature/s.

		A \Bitcoin \textbf{\tx} $\txid$ is identified by $\txidentifier$ computed as the hash of the \emph{body} of $\txid$,  i.e., $\hashfun(\txinput, \txoutput)$. $\txinput$ denotes the set of \addresses set as input and being spent in $\txid$; and $\txoutput$ is the set of addresses set as output. A transaction can have also a change output, where coins and address are owned by the same user controlling the inputs.

		We define a \Bitcoin \textbf{\entity} $\entityid$ as a set $\entityid := \set{\addressid_i}$ of \addresses controlled by the same user as clustered with the well-known and effective~\cite{harrigan2016unreasonable} co-spending heuristic~\cite{Meiklejohn2013}.
		This heuristic assumes that if two addresses (i.e. $\addressid_1$ and $\addressid_2$) are used as inputs in the same transaction while one of these addresses along with another address (i.e. $\addressid_2$ and $\addressid_3$) are used as inputs in another transaction, then the three addresses ($\addressid_1, \addressid_2, \addressid_3$) are likely controlled by the same actor.

        A \textbf{\Bitcoin wallet} is the software used by a \Bitcoin user to handle \Bitcoin \addresses owned by her.
        A wallet may correspond to a \Bitcoin entity, if addresses are reused.

	\subsection{Nodes and Payment Channels in the \LN (Layer 2)}

		A \textbf{\node} $\nodeid$ in the Lightning Network (\LN) is a tuple $\nodeid := (\nodeidentifier, \nodeip, \nodealias)$, where $\nodeidentifier$ is the identifier of the node; $\nodeip$ denotes the IP address associated with the \node, and $\nodealias$ the associated lexical label.

		A \textbf{payment channel} $\pcid$ is then created between two \nodes and denoted by the tuple $\pcid := (\chidentifier, \nodeid_1, \nodeid_2)$, where $\chidentifier$ denotes the channel's endpoint that is set to the identifier $\txid.\txidentifier$ of the \openchtx $\txid$ that created the channel. As the transaction may have several outputs, \chidentifier also contains the output index of the \channeladdress that locks the funds in the channel (e.g., \chidentifier:\channelOutputIndex); while $\nodeid_1$ and $\nodeid_2$ are the \nodes of the channel.

        An \textbf{\LN wallet} is the software used by an \LN user to manage her node, as well as the channels of this node. In practice, an \LN wallet comes with an integrated \Bitcoin wallet to open and close channels in the \LN. Recent releases of two \LN wallet implementations (\emph{lnd} and \emph{c-lightning})~\cite{lnd-release,c-lightning-release} enable opening/closing a channel using an external \Bitcoin wallet.

    \subsection{Cross-Layer Interaction}

		\begin{figure}[tb]
	\centering
	\includegraphics[width=\columnwidth]{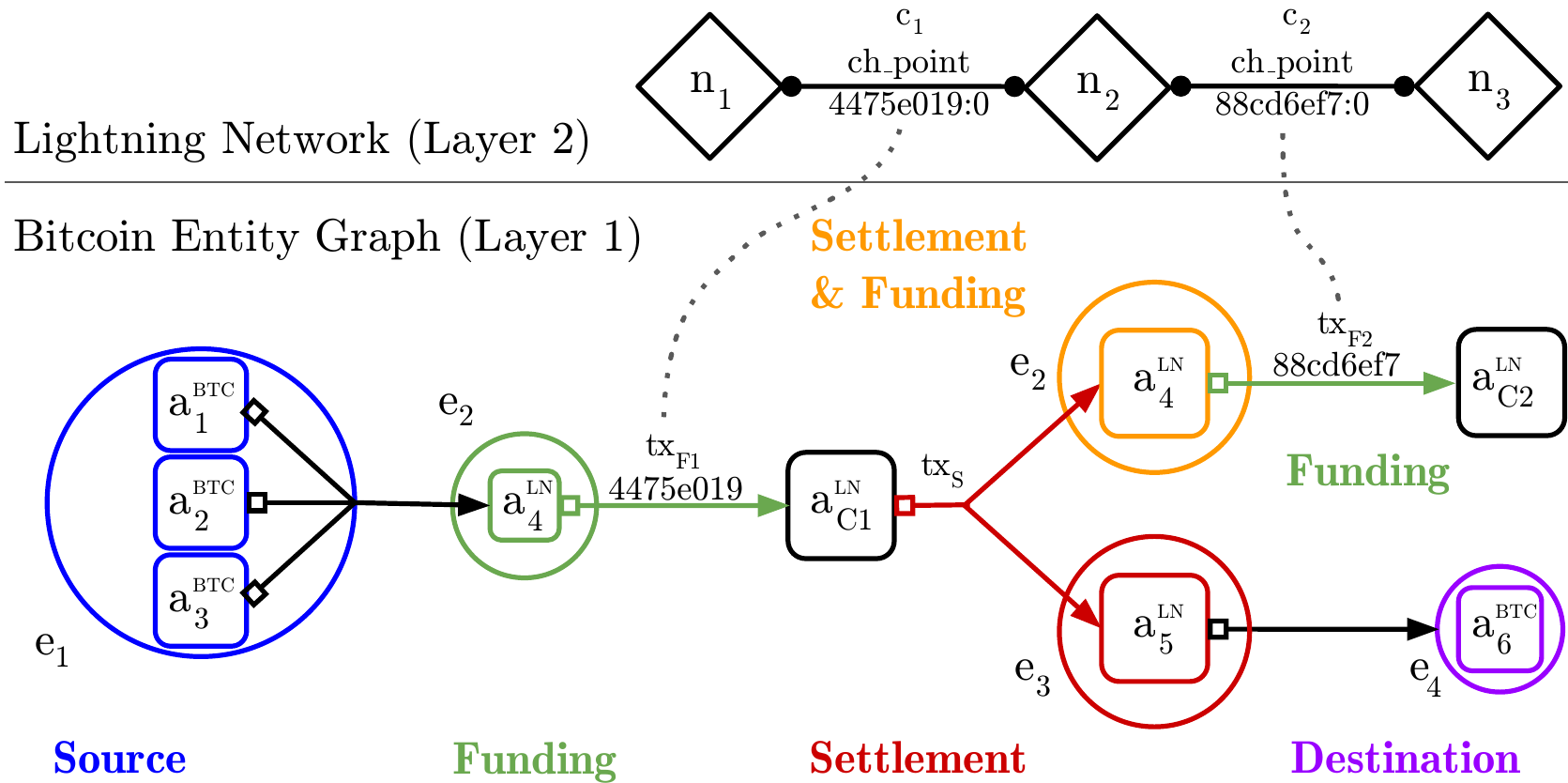}

	\caption{\textbf{Life cycle of an \LN channel.}
	At \lone, a source entity $e_1$ tops up entity $e_2$ that is then used in $tx_{F1}$ as funding entity of the channel $c_1$ represented by \channeladdress $a_{C1}^{LN}$.
	The channel $c_1$ is established at \ltwo between the nodes $n_1$ and $n_2$.
	The channel $c_1$ is then closed with the \closechtx $tx_S$ sending the funds back to two settlement entities, $e_2$ and $e_3$.
	The former, $e_2$, reuses these coins in $tx_{F2}$ to fund another channel ($c_2$) between $n_2$ and $n_3$ represented at \lone by the \channeladdress $a_{C2}^{LN}$.
	The coins in the other settlement entity, $e_3$, are instead collected into a destination entity $e_4$, not directly involved in the \LN.
	}\label{fig:coin_reuse_heuristic}
\end{figure}

		In this section, we describe the interaction between \Bitcoin and the \LN by means of the  example  illustrated in~\cref{fig:coin_reuse_heuristic}. Assume Alice wants to open a payment channel with Bob. Further, assume that Alice has a \Bitcoin wallet with coins in address $\addressid_1^{BTC}$ and she wants to open a payment channel with Bob. Additionally assume that Alice has never interacted with the \LN before and only has an \LN wallet, whose integrated \Bitcoin wallet handles $\addressid_4^{LN}$. In this setting, the lifetime of the payment channel between Alice and Bob is divided into the following phases:

        \paragraph{Replenishment}
        	Alice first transfers coins from her \Bitcoin wallet (represented by entity $e_1 := \set{\addressid_1^{BTC}, \addressid_2^{BTC}, \addressid_3^{BTC}}$) to her \LN wallet (entity $e_2 := \set{\addressid_4^{LN}}$), to top up the \LN wallet from the \Bitcoin wallet. We call $e_1$ the \textbf{source} entity as it is used as the source of funds to be later used in the \LN.

		\paragraph{Funding}
			Alice can now open a channel with Bob by first computing a \emph{deposit} \address $\addressid_{C1}^{LN}$ shared between Alice and Bob. In the next step, Alice creates a \textbf{\openchtx}  $tx_{F1}$ where $tx_{F1}.\txinput := \addressid_{4}^{BTC}$, $tx_{F1}.\txoutput := \addressid_{C1}^{LN}$, and $tx_{F1}.\txidentifier := \hashfun(tx_{F1}.\txinput, tx_{F1}.\txoutput)$.\footnote{Although theoretically a payment channel can be dual-funded (i.e., Bob also contributes $\balance_1$ to the \openchtx), this feature is under discussion in the community~\cite{dualFundedChannels2019} and currently only single-funded channels are implemented in practice.}
			After $tx_{F1}$ appears on the \Bitcoin blockchain, the payment channel $\pcid_1$ between Alice and Bob is effectively open. The channel $\pcid_1$ is then represented in the payment channel network as the tuple $(\pcid_1.\chidentifier, \nodeid_1, \nodeid_2)$, where $\nodeid_1$ and $\nodeid_2$ are \nodes belonging to Alice and Bob.

		\paragraph{Payment}
			After the channel $\pcid_1$ is open, during the \emph{payment} phase, both Alice and Bob can pay each other by exchanging authenticated \txs in a peer-to-peer manner authorizing the updates of the balance in the channel.
			Following our example, Alice and Bob create a \textbf{\closechtx} $\closechtxid$ where $\closechtxid.\txinput := \addressid_{C1}^{LN}$, $\closechtxid.\txoutput := \set{\addressid_4^{LN}, \addressid_{5}^{LN}}$ so that $\addressid_{4}^{LN}$ belongs to Alice, and $\addressid_{5}^{LN}$ belongs to Bob.
			The cornerstone of payment channels is that Alice and Bob do not publish $\closechtxid$ in the \Bitcoin blockchain. Instead, they keep it in their memory (i.e., off-chain) and locally update the balances in their channel $\pcid_1$. Both Alice and Bob can repeat this process several times to pay each other.

		\paragraph{Settlement}
			When the channel is no longer needed, Alice and Bob can close the channel by submitting the last agreed \closechtx into the \Bitcoin blockchain, thereby unlocking the coins from $\addressid_{C1}^{LN}$ into two \Bitcoin \addresses, each belonging to one of them with a number of coins equal to the last balance they agreed off-chain. In practice, the \closechtx may have more than two outputs: Alice can pay Bob to a third address where Bob needs to provide data other than a signature to redeem the coins (e.g., the valid preimage of a hash value before a certain timeout as defined in the Hash Time Lock Contract~\cite{htlc-wiki}).

		\paragraph{Collection}
			After the \closechtx appears in the \Bitcoin blockchain, Bob gets the coins in his \LN wallet. As a final step, Bob might want to get his coins into a different \Bitcoin wallet of his own. For that, Bob transfers funds from $\addressid_5^{LN}$ to $\addressid_6^{BTC}$, which we call \textbf{destination} address.

 		We note several points here. First, the addresses involved in the lifetime of payment channels could have been clustered into entities. In such a case, we refer to the source/funding/settlement/destination entity involved in the steps instead of the particular address itself.
  		In our example, Alice owns entity $\entityid_1$ that controls (among others) $\addressid_1^{BTC}$ and we thus say that entity $\entityid_1$ is the \emph{source}
 		entity in the replenishment step. Second, the same entity can be used at the same time for settlement and funding. Finally, Alice gets the coins from the channel with Bob in entity $\entityid_2$ that is then reused later to open a new payment channel.

	\subsection{The Cross-Layer Linking Problem}

		A starting point, as shown in~\cref{fig:coin_reuse_heuristic}, is to identify the \openchtx $\txid_{F1}$ corresponding to the payment channel $\pcid_1 := (\chidentifier, \nodeid_1, \nodeid_2)$, by finding the transaction (and the output index) that fulfills the condition $\txid_{F1}.\txidentifier = \pcid_1.\chidentifier$.
		While this is trivial, we cannot assert that the \entity $\entityid_2$ in $\txid_{F1}.\txinput$ also controls $\nodeid_1$, as it could also be that $\entityid_2$ controls $\nodeid_2$. Similarly, while we can deterministically get the \closechtx $\closechtxid$ used to close the channel $\pcid_1$, we cannot unambiguously link each settlement entity to the corresponding \node.

		The goal of this work is to cluster \Bitcoin entities based on their interactions with the \LN and then unambiguously link these clusters to \LN nodes that are under their control.
		Technically, this corresponds to finding a function that takes a set of \LN channels as input and returns tuples of the form (\entity, \node) for which it can be asserted that the \LN \node is controlled by the linked \Bitcoin \entity.

	\subsection{Related Work}\label{sec:related_work}
			Single-layer security attacks on the \LN topology were the focus of many recent studies: Rohrer et al.~\cite{Rohrer:2019aa} measured the \LN topology and found that the \LN is highly centralized and vulnerable to targeted (e.g., DoS) attacks.
			Similarly, Seres et al.~\cite{Seres:2020a} found that the \LN provides topological stability under random failures, but is structurally weak against rational adversaries targeting network hubs. Also, Martinazzi and Flori~\cite{Martinazzi:2020a} have shown that the \LN is resilient against random attacks, but very exposed to targeted attacks, e.g., against central players. Lin et al.~\cite{Lin:2020a} inspected the resilience of the \LN and showed that removing hubs leads to the collapse of the network into many components, evidence suggesting that this network may be a target for the so-called split attacks.
			Single-layer \LN privacy has recently been studied by Kappos et al.~\cite{Kappos:2020ab}, who focused on balance discovery and showed that an attacker running an active attack can easily infer the balance by running nodes and sending forged payments to target nodes.
			Nowostawski and Ton~\cite{mariusz_nowostawski_evaluating_2019} conducted an initial cross-layer analysis and investigated footprints of the \LN on the public \Bitcoin blockchain in order to find which transactions in the \Bitcoin blockchain are used to open and close \LN channels.
			Our work instead uses the funding and settlement transactions (and more) as input data to investigate for the first time: (i) the link between \LN nodes and \Bitcoin entities; (ii) clustering of \Bitcoin entities allowed by blockchain footprint for the interaction of these entities with the \LN; and (iii) the associated security and  privacy implications.


\section{Dataset}\label{sec:dataset}

In this section, we present the data we collected for our analysis.

    \subsection{Off-chain Data: \LN}
        We used the \LN Daemon (LND) software and captured  a copy of the \LN topology at regular intervals (30 min) via the \emph{describegraph} command since May 21 2019. The off-chain part of our dataset contains \nChannelsUsed channels, \nChannelsOpen of which were still open on September 9, 2020. The most recent channel in our dataset was opened on September 9, 2020, while the oldest was opened on January 12, 2018.
        We also define the \emph{activity period} of a node as the time that starts with the \openchtx that opened the first public channel in which the node appeared and ends either with the \closechtx of its last public channel or with 2020-09-09 (the time of preparing the dataset), if the nodes had still public channels open. Finally, we observe that channels in our dataset were established between \nNodesInChannels distinct nodes.

    \subsection{On-chain Data: \Bitcoin Blockchain}
        First, for each channel in our off-chain dataset, we used the \tx hash included in the channel's field $\chidentifier$ for retrieving the \emph{funding transaction}. Then, we checked whether the coins sent to the \channeladdress were spent or not. If a coin was spent, we fetched the \emph{settlement transaction}, that uses that \channeladdress as input. We obtained this data by querying the open-source GraphSense API\footnote{\url{https://api.graphsense.info/}} and the Blockstream API\footnote{\url{https://github.com/Blockstream/esplora/blob/master/API.md}}. We thereby extracted \nTxsFunding funding transactions\footnote{Some channels were opened with the same funding transaction.} and \nTxsClosing settlement transactions.
        Next, we extracted the input addresses of all funding transactions and the output addresses of all settlement transactions and mapped them to funding and settlement entities, as defined in~\cref{sec:background_blockchain}. Before clustering entities, we used BlockSci~\cite{Kalodner:2017a} to filter CoinJoin transactions because they would merge addresses of unrelated users. For the same reason, we also made sure that no CoinJoins from Wasabi nor Samourai\footnote{\url{https://github.com/nopara73/WasabiVsSamourai}} wallets were in our dataset. On the funding side, we also extracted the \emph{source entities} that were sending coins to funding entities; on the settlement side, we retrieved \emph{destination entities} that received coins from settlement entities. For that purpose, we implemented a dedicated data extraction and analytics job for the GraphSense Platform and executed it on a snapshot of the \Bitcoin blockchain up to block \nBlocks (2020-09-09 23:06), amounting for a total of \nTxsGS \txs and \nAddressesGS \addresses clustered into \nEntitiesGS \entities.
        After having extracted the \Bitcoin entities that were involved in opening and closing payment channels, we attributed them using the Chainalysis API\footnote{\url{https://www.chainalysis.com/}} and assigned service categories (e.g., exchange, hosted wallet) to entities.

        \begin{table}[tb]
            \centering%
            \caption{On-chain Dataset Summary.}
            \adjustbox{max width=\columnwidth}{
\begin{tabular}{l r r r r}
	\toprule
	& Source & Funding & Settlement & Destination\\
	\midrule
\rowcolor{Gray}\# Addr &  & 170,777 & 88,166 & \\
\# Entities &  196,131 & 96,838 &  53,371 & 424,732\\
\rowcolor{Gray}\# Addr (Exp.) & 70,638,581 & 196,818 & 2,243,525 & 107,474,279\\
\# Services & 5,812 & 1 & 5 & 67,969\\
\rowcolor{Gray}\# Relations & & 203,328 & 438,725 & 	\\
\bottomrule
\end{tabular}
}
            \label{tab:dataset_overview}
        \end{table}

        Table~\ref{tab:dataset_overview} summarizes the number of addresses (\emph{\# Addr}) found in funding and settlement transactions as well as the number of resulting entities after applying the co-spending heuristic on these addresses (\emph{\# Entities}). We can clearly observe that the number of distinct source entities (\nEntitiesSource) is lower than the number of destination entities (\nEntitiesDestination), which is also reflected in the number of relations (\emph{\# Relations}) representing monetary flows from source to funding entities and from settlement to destination entities, respectively. These unbalanced numbers might be due to funds going from settlement entities to mixing services, as we discuss later.
        Since the co-spending heuristic also groups addresses which were not part of our dataset snapshot, we also added the number of expanded addresses (\emph{\# Addr (Exp.)}). The difference between the number of addresses and entities on both the source and destination side can be explained by the presence of super-clusters, which are responsible for large transaction inputs and outputs and typically represent service entities such as cryptocurrency exchanges~\cite{harrigan2016unreasonable}.
        Finally, this table also lists the number of identified service entities (\emph{\# Services}). We only found them in few cases for funding (1) and settlement (5) entities, probably because mostly non-custodial wallets are used when opening and closing channels and known services in our dataset behave only as source and destination entities. Roughly 0.9\% of all source entities were categorized, with the majority (0.8\%) being exchanges. On the settlement side, we identified 10\% of all destination entities as wallets being controlled by services, with the majority (8\%) being mixing services. We can not fully account for this strong connection to mixing, but it does suggest that many \LN users are privacy-aware. Indeed, there is evidence that the \LN is recognized as a privacy technology complementary to mixing. e.g., the well-known mixing wallet Wasabi suggests \LN as one way to enhance privacy when using the wallet\footnote{https://docs.wasabiwallet.io/using-wasabi}.

    \subsection{Ground Truth Data: \LN Payments}
        We devised and implemented a simple process that allows us to create a ground truth dataset of entity-node pairs that can then be compared with our linking results as a validation step. We first run our \algorithmslink resulting in an initial set of entity-node pairs. We then found a trade-off for selecting the target nodes: some randomly-selected linked nodes for generality purposes and some other nodes with the highest number of \closechtxs as a sign of being very active on the network and reusing funds, a useful aspect for the next steps. Next, we managed to open channels, perform payments and close channels with \nNodesReceivedOurCoins of them. For these nodes that received coins from us, we are able to see their settlement entity, but only \nNodesReceivedSpentOurCoins nodes further spent the settlement funds in other transactions, necessary for us to capture their spending behaviors with our heuristics.
        We additionally managed to have channels open to us from \nNodesOpenedChannelsToUs \LN nodes that provide inbound channels as a service, revealing their funding entities. We performed this activity at the beginning of September 2020 (block 646559) and after waiting some days to let the nodes spend our coins, we run the \algorithmslink again on our latest dataset (until block \nBlocks) so that for these targeted nodes we have both ground truth and heuristically-obtained links to entities. In~\cref{sec:validation}, we compare this ground truth data with our linking results, while a more detailed explanation of the methodology to extract this data is presented in~\cref{app:ground-truth}.


\section{Clustering Heuristics}\label{sec:clustering}
    In this section we introduce the on-chain and off-chain clustering heuristics.

    \subsection{On-Chain \Bitcoin Entity Clustering (Layer 1)}\label{sec:clustering_on_chain}

\begin{figure}[htbp]
            \centering
            \includegraphics[width=.9\columnwidth]{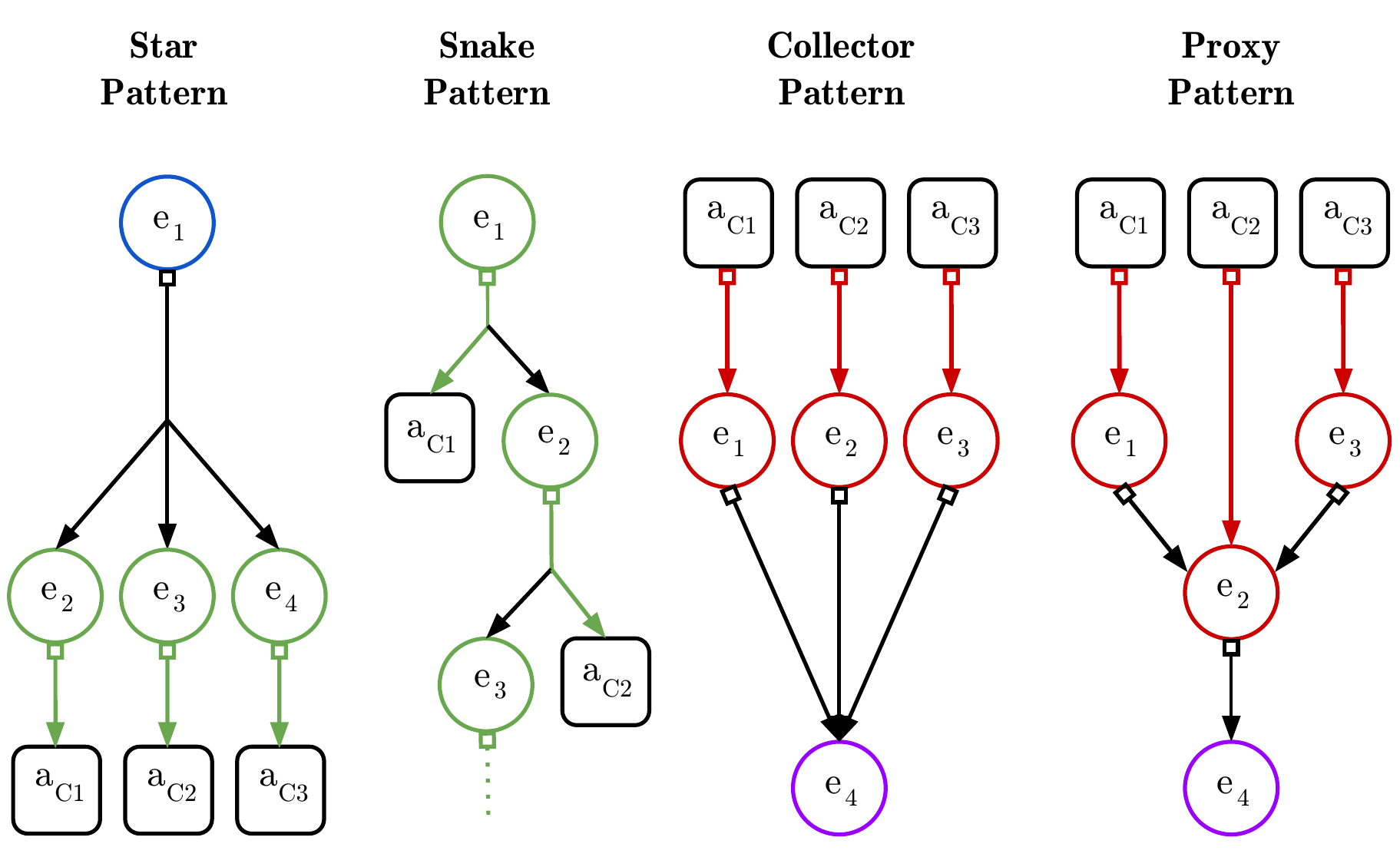}
            \caption{On-chain clustering heuristics. Following the same notation of~\cref{fig:coin_reuse_heuristic}, in the star pattern, a source entity $e_1$ replenishes  three different funding entities creating a single cluster ($e_1$, $e_2$, $e_3$, $e_4$). In the snake pattern, a series of funding transactions are performed using the change address of a previous funding transaction as input and the funding entities can be clustered ($e_1$, $e_2$, $e_3$). In the collector and proxy pattern, multiple settlement entities merge their coins in one single entity and these settlement entities can be clustered ($e_1$, $e_2$, $e_3$, $e_4$).}\label{fig:on-chain-example}
        \end{figure}

        \LN-blockchain interactions result in monetary flows from source to funding and from destination to settlement entities (see~\cref{fig:coin_reuse_heuristic}). When analyzing the resulting entity graph abstraction, we observed four patterns (see \cref{fig:on-chain-example}).

        First, several funding entities received funds from the same source entities with one source entity transferring coins to several funding entities. This forms a \emph{star-shaped pattern} and reflects a current \LN wallet feature, which requires\footnote{We note that this requirement may no longer be there if the "fund-from-external-wallet" functionality, already available in the recent release~\cite{lnd-release}, is widely adopted.} users to transfer funds from an external wallet (source entity) to an internal wallet (funding entity) before opening a channel. If these source entities are not services, which is rarely the case (see \cref{sec:dataset}), then we can assume:

        \begin{definition}[Star Heuristic]
        If a component contains one source entity that forwards funds to one or more funding entities, then these funding entities are likely controlled by the same user.
        \end{definition}

        Second, again on the funding side, we observed a \emph{snake-like pattern} in which source entities transfer coins to a funding entity, which then opens a channel and the change from the funding transaction is used to fund another channel, and so on (analogous to the Bitcoin Change-Heuristic~\cite{Meiklejohn2013}).

        \begin{definition}[Snake Heuristic]
        If a component contains one source entity that forwards funds to one or more entities, which themselves are used as source and funding entities, then all these entities are likely controlled by the same user.
        \end{definition}

        Third, we identified a so-called \emph{collector pattern}, which mirrors the previously described star pattern on the settlement side: a user forwards funds from several settlement entities, which hold the unlocked coins of closed channels in an internal wallet, to the same \emph{destination entity}, which serves as an external \emph{collector} wallet of funds and therefore fulfills a convenience function for the user.

        \begin{definition}[Collector Heuristic]
        If a component contains one destination entity that receives funds from one or more settlement entities, then these settlement entities are likely controlled by the same user.
        \end{definition}

        Fourth, we found a refined collector pattern, which we call \emph{proxy pattern}: a user first aggregates funds from several settlement transactions in a single settlement entity and then forwards them to a single destination entity.

        \begin{definition}[Proxy Heuristic]
        If a component contains one destination entity that receives funds from one or more entities, which themselves are used as settlement and destination entities, then these entities are likely controlled by the same user.
        \end{definition}

        We compute the above heuristics as follows: we construct 1-hop ego-networks for the funding and settlement entities and extract funding relations and settlement relations (see~\cref{sec:dataset}). Next, we compute all weakly-connected components in these graphs and filter them by the conditions defined above.

        \begin{table}[tb]
            \centering%
            \caption{On-chain clustering results.}\label{tab:clustering_results_onchain}
            \adjustbox{max width=\columnwidth}{
	\begin{tabular}{l l r r r r}
		\toprule
		         		& Star (F) 	& Snake (F) 	& Collector (S) & Proxy (S)\\
		\midrule
		\rowcolor{Gray}\# Components  	& \nStars (\percStars\%)	& \nSnakes (\percSnakes\%) 	& \nCollectors (\percCollectors\%)	& \nProxies (\percProxies\%)\\
		\# Entities 	& \nEntitiesStar & \nEntitiesSnake (\percEntitiesSnake\%)	& \nEntitiesCollector (\percEntitiesCollector\%)	& \nEntitiesProxy (\percEntitiesProxy\%)\\
		\rowcolor{Gray}\# Addresses 	& \nAddressesStar & \nAddressesSnake & \nAddressesCollector 	& \nAddressesProxy \\
		\bottomrule
	\end{tabular}
}
        \end{table}

        \cref{tab:clustering_results_onchain} shows the number of \Bitcoin entities we were able to cluster with each heuristic. When regarding the connected components, we can clearly see the rare occurrence of the star patterns and the dominance of the snake pattern, which represents \percSnakes\% of all funding components. On the settlement side, \percCollectorsProxies\% of all components either match the collector or the proxy pattern. Consequently, we were able to group \percEntitiesFundingClustered\% (\nEntitiesFundingClustered) of all funding entities and \percEntitiesSettlementClustered\% (\nEntitiesSettlementClustered) of all settlement entities. This corresponds to \nAddressesFundingClustered funding addresses and \nAddressesSettlementClustered settlement addresses.

        \paragraph{Discussion}
            Our heuristic can, by definition, also yield false positives for two main reasons: first, an entity could represent several users if clustered addresses are controlled by a service (e.g., exchange) on behalf of their users (custodial wallet) or if transactions of several unrelated users are combined in a CoinJoin transaction. Second, users could transfer ownership of \Bitcoin wallets off-chain, e.g., by passing a paper wallet. While the second case is hard to filter automatically, we applied countermeasures to the first case: first, we filtered known CoinJoin transactions (see \cref{sec:dataset}), and second, we filtered all components containing service entities by using Chainalysis, one of the most comprehensive attribution dataset available.

        \paragraph{Countermeasures}
            We suspect that the above patterns reflect a user behavior that is already known to compromise the privacy of transactions: reuse of TXOs (transaction outputs). If outputs of funding transactions are not reused for opening other channels, the snake heuristic would not work; if users refrain from funding channels from a single external source and avoid collecting funds in a single external destination entity, the other heuristics would not yield any significant results. Despite not pervasive on the network, Coinjoins and similar solutions could, in theory (e.g., if used as funding transactions), obfuscate the entity linked to an \LN node behind a set of unrelated addresses.

    \subsection{Off-Chain \LN Nodes Clustering (Layer 2)}

        We have also designed an algorithm to cluster \LN nodes based on aliases and IPs reported in the \LN, along with their corresponding autonomous systems (AS). If a set of node aliases share a common substring, and they are hosted on the same AS, we cluster them. Similarly, if a set of nodes report the same IP or onion address, we cluster them assuming they are controlled by the same entity. This allows us to cluster $1,251$ nodes into $301$ clusters. Due to space constraints, we defer the description of this clustering to~\cref{sec:clustering_off_chain}.

\section{Linking \LN Nodes and \Bitcoin Entities}\label{sec:linking}

	In this section, we present two algorithms that link \LN \nodes to the \Bitcoin \entities that control them.
	In both of these heuristics, we do not consider \closechtxs with more than two output entities (\percTxSettlementMoreThanTwoOutputs\% of the \closechtxs), as they are not a cooperative close and do not allow us to unambiguously link nodes and output entities. Furthermore, we ignore \closechtxs that involve punishment transactions~\cite{bolt3}. Finally, an assumption that we make in both of the following \algorithmslink is that if one node in a channel has been linked to a settlement entity and the \closechtx has two output entities, then the other node can be linked to the other settlement entity.

	\subsection{Linking Algorithm 1: Coin Reuse}\label{sec:heuristic1}

		Our \algorithmlink builds upon the usage pattern that appears when a payment channel is closed and the user that receives the coins from such channel reuses them to open a new payment channel.
		An illustrative example of this \algorithmlink is included in~\cref{fig:coin_reuse_heuristic} where a funding \entity $\entityid_2$ has been used to open a channel $\pcid_1$ between \nodes $\nodeid_1$ and $\nodeid_2$ with the \openchtx $\txid_{F1}$.
		Later, this channel has been closed in the \closechtx $\txid_S$, releasing the coins in the channel to the \entities $\entityid_2$ (i.e., the same that was used as input in $\txid_{F1}$) and $\entityid_3$. Finally, assume that the owner of \entity $\entityid_2$ decides to open a new channel reusing the coins from  $\txid_S$ performing a new \openchtx $\txid_{F2}$ which results in the payment channel $\pcid_2$ between the aforementioned \node $\nodeid_2$ and  $\nodeid_3$.
		In this situation, given that the entity $\entityid_2$ has appeared in the \closechtx of $\pcid_1$ and has been reused to open a new channel in the \openchtx $\pcid_2$,  our heuristic concludes that the entity $\entityid_2$ controls node $\nodeid_2$.

		\begin{definition}[\Algorithmlink 1: Coin Reuse]
			Assume that a \Bitcoin entity $\entityid$ opens an \LN channel $\pcid_1 := (\chidentifier_1, \nodeid_1, \nodeid_2)$.
			If $e$ is used as settlement entity to close the \LN channel $\pcid_1$ and also as funding entity to open a new \LN channel $\pcid_2 := (\chidentifier_2, \nodeid_1, \nodeid_3)$, and the nodes $\nodeid_2$ and $\nodeid_3$ have activity period overlap, then the user controlling entity $\entityid$ also controls the \LN node $\nodeid_1$ in common to both channels $\pcid_1$ and $\pcid_2$.
		\end{definition}

		We applied the \algorithmlink based on coin reuse which resulted in \nLinksReusedHeurOne tuples of (\openchtx, \closechtx, \openchtx) and \nEntitiesReusedHeurOneNone \entities reusing their addresses for opening and closing channels.
		Once these \nEntitiesReusedHeurOneNone entities are linked to \LN nodes, all the other output entities in the \closechtxs of these \nEntitiesReusedHeurOneNone entities can be linked to the counter-party nodes in the channels as mentioned earlier. Finally, after these new links are created, our heuristic can iteratively go over the settlement transactions that involve these newly linked entities to find other entity-node pairs.

		After 7 iterations, the heuristic yielded \nEntitiesLinkedHeurOneNone entities linked to \nNodesLinkedHeurOneNone nodes, thus having cases where a node is linked to multiple entities. In total, considering the number of entities we have in our dataset (\nEntitiesFundingSettling overall, both funding and settlement side\footnote{Here we do not consider source and destination entities as they do not directly interact with the \LN.})
		the heuristic is able to link \percEntitiesLinkedHeurOneNone\% of them. This result is a lower bound on the possible number of linked entity-node pairs because the \algorithmlink mainly relies on channels to be closed (in our dataset only half of them are) and on a specific subset of entities, namely the output entities of settlement transactions with exactly two outputs, one per node. In fact, if we focus only on settlement transactions with two output entities, we have \nEntitiesTwoOutputsNone entities, \percEntitiesLinkedHeurOneTwoOutputsNone\% of which can be linked, showing thereby that this \algorithmlink has a targeted but effective linking effect.
		Regarding the nodes percentages, we can link \percNodesLinkedHeurOneNone\% of the total (\nNodesInChannels overall) and \percNodesLinkedHeurOneTwoOutputsNone\% of the nodes for which there exists at least one channel that has been closed using a 2-output-entity settlement transaction, confirming the trend we observed with entities.

		\paragraph{Discussion}\label{sec:linking1_validation}
			We note that requiring that the same \entity is used for all three transactions (i.e., funding and settlement of the first channel as well as funding of the second channel) may be too restrictive and leave out further links of entities and nodes. However, we enforce this restriction  to avoid false positives that could be otherwise introduced as we describe next.
			Assume we control an \LN node, $n_2$, with an associated \Bitcoin entity $e_1$ that funds channel $\pcid_1$ between node $n_2$ and $n_1$ through $tx_{F1}$. Furthermore, we have an \LN wallet with an associated \Bitcoin entity, $e_3$, on our phone provided by a third-party app. This means that there must be another node in the \LN, $n_3$, managed by this third-party app. When we decide to close channel $\pcid_1$, we specify an address provided by our third-party app, hence belonging to entity $e_3$, as settlement address to receive the funds back. We finally proceed to use these funds to open a new channel, $\pcid_2$, again with node $n_1$ but from node $n_3$, the third-party node. Without the requirement on the same funding entity, the heuristic would link the node $n_1$, in common between the two channels, to the entity $e_3$ reusing the funds, which is false. With the same funding address requirement, instead, this case is ignored.
			A further condition that needs to be satisfied to strengthen this heuristic is that the nodes not common to the two channels (nodes $n_1$ and $n_3$ in~\cref{fig:coin_reuse_heuristic}) have a time overlap in their activity period. This excludes the unlikely, but not impossible case that one node changes its ID (public key) from $n_2$ to $n_3$ keeping the same \Bitcoin wallet (and thus entity), which could allow one to open two channels from two different nodes, but to the same node, using the same \Bitcoin entity, creating a false-positive case for the heuristic.

        \paragraph{Countermeasures}
        	The default functionality of \LN wallets followed thus far by virtually all users consists of having a single wallet per node from where to extract the funds to open channels and where to send the coins after channels connected to such node are closed. We conjecture that this setting favors the usage pattern leveraged in the \algorithmlink described in this section. As a countermeasure, we advocate for the support of funding and settlement channels of a single node from different (external) \Bitcoin wallets, helping thus to diversify the source of funds. We observe that recent versions of the \LN wallet \emph{lnd} and \emph{c-lightning} have started to support this functionality~\cite{lnd-release,c-lightning-release}.

	\subsection{Linking Algorithm 2: Entity Reuse \label{sec:heuristic2}}

		In this \algorithmlink we leverage the usage pattern that appears when a user reuses the same \Bitcoin wallet (e.g., the one integrated within the \LN wallet) to open several payment channels. Thus, in this \algorithmlink we assume that an \entity $\entityid$ opened several payment channels with other \entities. This common usage pattern in practice can be detected at the blockchain by finding the set of $\numberofchannels$ \openchtxs that have $\entityid$ in common as the funding \entity. We can thus say that $\entityid$ has opened  $\numberofchannels$ channels. At the \LN, if there is only one \node $\nodeid$ common to all the $\numberofchannels$ channels funded by $\entityid$, we say that $\entityid$ controls $\nodeid$. An illustrative example of this \algorithmlink is shown in~\cref{sec:examplelinking}.

        \begin{definition}[\Algorithmlink 2: Entity Reuse]
        	If there are $\numberofchannels > 1$ channels opened by one single funding entity $\entityid$ that have only one \LN node $\nodeid$ in common, and there are at least two nodes $\nodeid_x$ and $\nodeid_y$ in these channels with activity period overlap, then the user controlling entity $\entityid$  controls  node $\nodeid$ too.
        \end{definition}

		We can link \nEntitiesLinkedHeurTwoMinConfNone \entities to \nNodesLinkedHeurTwoMinConfNone nodes which correspond to \percEntitiesLinkedHeurTwoMinConfNone\% of all the entities and \percNodesLinkedHeurTwoMinConfNone\% of all the nodes respectively.

		\paragraph{Discussion}\label{sec:linking2_validation}
			The way this \algorithmlink has been described and implemented so far might yield false entity-node links. As discussed in section~\ref{sec:linking1_validation}, a user can open a channel from its node $n_2$ to another node $n_1$, then close the channel, change its node ID to $n_3$ keeping the same \Bitcoin wallet and finally open a second channel to $n_1$.
			For this \algorithmlink, this example would cause a false positive because $\nodeid_1$ would be linked to the \Bitcoin entity of this user. To prevent this from happening, we add the following condition. Consider the set of nodes appearing in the channels funded by a single \fundingentity $e$ and exclude from this set the node that has been linked to $e$ with this heuristic. Now, if there is at least one pair of nodes ($\nodeid_2$, $\nodeid_3$ from the example above) in this set that have an activity period overlap, then we discard the false-positive risk as it is not possible for node $n_2$ to change to $n_3$ keeping two channels open.
			When implementing this additional requirement, we discovered that our results do not contain any false positive as there is at least one pair of nodes with an activity period overlap for each entity-node link.
			To further validate the results of this second \algorithmlink, we report that it provides the same entity-node links that are in common with the \algorithmlink presented in~\cref{sec:heuristic1}.

        \paragraph{Countermeasures}
           	A countermeasure to this heuristic is to not reuse the same funding entity to open multiple channels. This can be achieved either by having multiple unclustered addresses in a wallet or to rely on external wallets~\cite{lnd-release,c-lightning-release}.

    \subsection{Validation}\label{sec:validation}

		For the validation of the heuristics presented in this work we use the ground truth dataset presented in~\cref{sec:dataset}. For each of the \nNodesReceivedSpentOurCoins nodes that received funds from us, we compare their set of ground truth settlement entities with their set of linked entities from our \algorithmslink. If there is an intersection between these two sets, we say that the link is validated. In total, we find that \nNodesReceivedSpentOurCoinsValidates  nodes (i.e., $63\%$) are validated.
		The validation for the \nNodesOpenedChannelsToUs nodes that opened channels to us is the same, but uses their ground truth funding entities as set for comparison with the set of linked entities from our \algorithmslink. In this case, we can validate \nNodesOpenedChannelsToUsValidated nodes.
		The lack of validation for the other nodes can have several reasons: i) as reported in~\cref{sec:dataset}, we notice that only $11$ out of the $52$ nodes receiving our coins (by default on newly-generated \Bitcoin addresses) also spent them, ii) the coins spent are not merged with funds from other channels or iii) the coins are spent and merged with funds from channels missing in our dataset. Nevertheless, one should note that over time our ground truth data will increase and more nodes could be validated as soon as they spend our funds.

		We believe that our small ground truth dataset is a reasonable trade-off between obtaining a representative picture of the \LN main net and a responsible and ethical behavior that does not alter the \LN properties significantly. We also see our methodology to gather ground truth data as an interesting contribution due to its scalability features: costs are relatively low (two on-chain transactions and \LN routing fees for each targeted node) and executable in a programmatic way. We defer a more detailed description of this methodology to~\cref{app:ground-truth}.


\section{Assessing Security and Privacy Impact}\label{sec:combined}

    We merged the results of our \algorithmscluster (\cref{sec:clustering}) and our \algorithmslink (\cref{sec:linking}), thereby increasing the linking between entities and nodes as shown in~\cref{tab:linking_clustering_results_diff_summary}. We defer to~\cref{app:heuristics-contribution} a detailed description of the contribution for each heuristic individually.

    \subsection{Privacy Impact on \Bitcoin Entities (Layer 1)}
        The \algorithmslink and \algorithmscluster described in this work allow attributing activity to \Bitcoin entities derived from their interaction with the \LN. Assume that a cluster is formed by a certain number of \Bitcoin entities and \LN nodes, then if any of the \LN nodes has publicly identifiable information (e.g., alias or IP address), this information can be attributed to the \Bitcoin entities as well.   In total, we can attribute tagging information to \nEntitiesAliasIPTagged different entities that in total account for \nAddressesAliasIPTagged different addresses, which represent \percAddressesAliasIPTagged\% of our dataset.

        This deanonymization is based purely on publicly available data\footnote{We note that Chainalysis attribution data is not strictly necessary for the \algorithmslink.} and can be carried out by a low budget, passive adversary that simply downloads the \Bitcoin blockchain and the information from the \LN. We envision that further impact can be achieved by a more powerful adversary (e.g., a \Bitcoin miner).
        Moreover, the possible deanonymization of \Bitcoin entities hereby presented shows that it is crucial to consider the privacy of both layers simultaneously instead of one of them at a time as largely done so far in the literature.

    \subsection{Security and Privacy Impact on the \LN (Layer 2)} We have evaluated the implications of our clustering and linking algorithms in the  security and privacy of the \LN. In summary, we studied how the capacity of the \LN is distributed across actors and found that a single actor controls over $24\%$ of the total \LN capacity and as few as five actors consisting of 36 nodes control over $33\%$ of the total capacity. Few \LN actors are thus in a privileged situation that can be used to diminish the security and privacy of the \LN. For instance, we observed that the entity with the highest capacity can render useless over $40\%$ of the channels for a period of time by means of DoS attacks. Similar issues appear from the privacy point of view, where just 5 actors can learn the payment amounts used in up to 60\% of the cheapest paths in the \LN and determine who pays to whom in up to 16\% of the cheapest paths. Due to space constraints, we defer a detailed discussion of our security and privacy assessment to~\cref{app:secanalysis}. 

	\begin{table}[tbp]
		\centering
		\caption{Summary results }\label{tab:linking_clustering_results_diff_summary}
		\adjustbox{max width=\columnwidth}{%
		\begin{tabular}{@{}lrrr@{}}
			\toprule
			Linking + Clustering                         & \multicolumn{1}{c}{\begin{tabular}[c]{@{}c@{}}\% addresses\\ linked\end{tabular}} & \multicolumn{1}{c}{\begin{tabular}[c]{@{}c@{}}\% entities\\ linked\end{tabular}} & \multicolumn{1}{c}{\begin{tabular}[c]{@{}c@{}}\% nodes\\ linked\end{tabular}} \\ \midrule
		    \rowcolor{Gray} \Algorithmlink 1 & \percAddressesLinkedHeurOneNone & \percEntitiesLinkedHeurOneNone & \percNodesLinkedHeurOneNone \\
			\Algorithmlink 1 + all on/off-chain & \percAddressesLinkedHeurOneAll & \percEntitiesLinkedHeurOneAll & \percNodesLinkedHeurOneAll \\
			\rowcolor{Gray} \Algorithmlink 2 & \percAddressesLinkedHeurTwoMinConfNone & \percEntitiesLinkedHeurTwoMinConfNone & \percNodesLinkedHeurTwoMinConfNone \\
			\textbf{\Algorithmlink 2 + all on/off chain} & \percAddressesLinkedHeurTwoMinConfAll & \percEntitiesLinkedHeurTwoMinConfAll & \percNodesLinkedHeurTwoMinConfAllAliasIP\\
			\bottomrule
		\end{tabular}
		}
	\end{table}

\section{Conclusion and Future Work \label{sec:conclusion}}
	In this paper, we presented two novel \algorithmslink to reveal the ownership of \Bitcoin addresses that are controlled by \LN nodes using publicly-available data. We also developed four \Bitcoin address \algorithmscluster and one \LN node \algorithmcluster that allowed us to link \percAddressesLinkedHeurTwoMinConfAll\% of the \Bitcoin addresses in our dataset to \percNodesLinkedHeurTwoMinConfAllAliasIP\% of the public \LN nodes, and cluster \nNodesAliasIPCluster \LN nodes into \nClustersOffChainOnly actors. Finally, we discussed the security and privacy implications of our findings in the \LN, where we find that a single actor controls 24\% of the overall capacity and a few actors have a large impact on value privacy and payment relationship anonymity. These few actors also have a large overlap with those  that would be candidates for high-impact attacks, the success of which can have significant negative effects on payment success and throughput for the entire \LN. 

	Scalability issues appear in a broad range of blockchain applications and \lMinusTwo protocols are increasingly considered as possible solutions. In light of these developments, we find an interesting venue for future work to evaluate whether our heuristics apply to \lMinusTwo protocols other than the \LN such as the Raiden Network for Ethereum.

\ifanonymous
\else
	\section*{Acknowledgments}
		This work is partially funded by the European Research Council (ERC) under the European Unions Horizon 2020 research (grant agreement No 771527-BROWSEC), by PROFET (grant agreement P31621), by the Austrian Research Promotion Agency through the Bridge-1 project PR4DLT (grant agreement 13808694); by COMET K1 SBA, ABC, by CoBloX Labs, by the Austrian Science Fund (FWF) through the Meitner program (project agreement M 2608-G27) and by the Austrian security research programme KIRAS of the Federal Ministry of Agriculture, Regions and Tourism (BMLRT) under the project KRYPTOMONITOR (879686).
		The authors would also like to thank Peter Holzer and Marcel M\"{u}ller for setting up and starting the \LN data collection process.
\fi

\bibliographystyle{splncs04}
\bibliography{bibliography}

\appendix

\section{Ground Truth Data Collection\label{app:ground-truth}}

		To send coins to \LN nodes and discover their settlement entities, we run two nodes in the \LN. One is the \emph{sending} node, $n_s$, and the other the \emph{receiving} node, $n_r$. We make sure that $n_r$ is connected to the \LN by a channel with a good amount of incoming capacity, so that it can receive a number of \LN payments. We then have $n_s$ open a channel, $c_t$, to a given \emph{target} node, $n_t$. Once $c_t$ is open, we route a payment of amount $a$ from $n_s$ to $n_r$ over $c_t$. On successful payment, we close the channel $c_t$. We then repeat this experiment for a number of target nodes $n_t$.

		The purpose of making a payment in this way is to increase the balance of $n_t$
		on channel $c_t$ to $a$ before the channel is closed. This ensures that when
		$c_t$ is closed, the entity of $n_t$ receives an on-chain payment of $a$ BTC. If
		the entity further spends this amount of BTC we can apply our heuristics and
		attempt to link entity to \LN node. An advantage of routing the payment over
		$n_t$ rather than making the payment to $n_t$ directly, is that we will not have
		to request an invoice from each $n_t$, or rely on the currently experimental
		\emph{key send} mechanism. Instead, we just have $n_r$ generate a set of invoices
		we can use for the experiment.

		For each target node, we attempted to open a
		channel of 100,000 satoshis capacity and make a payment of 1,000 satoshis to
		the receiving node. We chose these amounts as they allow us to perform the
		experiments with a relatively low amount of capital. While the payment is low,
		it is above the \Bitcoin dust limit of 546 satoshis, ensuring that the target node
		will receive the funds.

		Unfortunately, in some cases the experiment would fail, either because the
		channel would fail to open, or the payment was unsuccessful. When the channel
		fails to open most often this is due to the target node not responding to the
		channel opening request (presumably as the node is no longer
		on-line). Occasionally, our requests to open channels will fail due to the
		requested channel not meeting a policy set by the target node for opening new
		channels (e.g., some nodes will only accept channels above a certain
		capacity). Once a channel was established, the payment could fail because a
		suitable route could not be found between receiving and sending node, although
		this was rare.

\section{Off-Chain \LN Node Clustering}\label{sec:clustering_off_chain}
    The operator of an \LN node can announce custom node features such as an alias, which was added to the \LN to improve the usability of the system.
    The alias can be changed by the operator at any time without affecting the operation of open channels, as those are only tied to a node's private and public key pair.
    We observed that when a user is operating multiple nodes, it is likely that she will name her nodes in a similar fashion, or along a common theme. For example, the operator LNBIG.com enumerates its nodes on their website\footnote{\url{https://lnbig.com/#/our-nodes}}, with aliases such as LNBIG.com [lnd-25], LNBIG.com [lnd-34]. Via public chat, the developers confirmed that LNBIG.com Billing also belongs to them. Strong alias similarities are most likely intentional, for example, to make it easier for users to identify a service, or the operators may want to achieve a reputation or branding effect.

    In order to find nodes under the control of the same entity, we can exploit the alias information and measure similarity.
    We evaluated popular string similarity metrics (cf.~\cite{gomaa2013survey}) such as the Levenshtein, Hamming and Jaro-Winkler distances. Naturally, however, aliases can be similar, but do not belong to the same entity. Examples include node aliases such as WilderLightning and GopherLightning, which overlap textually but are not controlled by the same entity.

    Apart from the alias, nodes advertise their IP address (or an address within the Tor network) and a port. We can use this additional public information to filter the clusters obtained through alias similarity, increasing the confidence that the nodes are operated by the same entity.

    Each IP address is part of a Classless Inter-Domain Routing (CIDR)~\cite{fuller1993classless} prefix that is under the control of one or multiple network operators. An Internet Service Provider (ISP) may operate a collection of such CIDRs, and their grouping is called autonomous system (AS), each of which is identified by an autonomous system number (ASN). By performing WHOIS queries, we can obtain the ASN's of each \LN node IP address. If an alias-based node cluster consists only of IP addresses associated to a single ASN, we conclude that the \LN nodes are hosted by the same network operator and is, therefore, more likely operated by a single entity. In addition, we also cluster \LN nodes that are (or have been in the past) reachable via the same IP or Tor address.

    Technically, we first determine \emph{pairwise alias distances} by computing a distance matrix between all \LN node aliases using different distance metrics. Then we perform \emph{agglomerative hierarchical clustering} to avoid early cluster merging due to single aliases being similar to two distinct clusters. For \emph{threshold identification}, we evaluate the full range of thresholds by counting the number of \LN nodes that remain when pruning clusters that are not pure with respect to their ASNs. We then choose the threshold that results in the largest number of clustered nodes, while ensuring the LNBIG.com cluster is identified as a single cluster of at least 26 nodes, as we have ground truth from their website. In parallel, we perform \emph{IP-based clustering} by grouping all \LN nodes that have been seen to be reachable via the same IP or Tor address. Finally, we \emph{join alias and IP-based clustering} and merge the resulting alias and IP-based clusters if there is an overlap. This results in the final off-chain-based \LN node clusters.

    In our analysis, we considered all nodes with their history of aliases and  valid addresses. IPs within address ranges reserved for special purposes such as private networks are excluded. We compared the performance of ten different string distance measures (see~\cref{sec:evaluation-string-distance}) and concluded that the relative longest common substring measure yields the best results. In particular, 363 \LN nodes have been grouped into 126 clusters. The IP-based clustering yields 1135 clustered \LN nodes, 241 of which are already part of the alias-based clusters. By merging these clusters, the final cluster count is \nClustersOffChainOnly, with a total of \nNodesAliasIPCluster \LN nodes clustered. The two largest clusters are nodl-lnd-s007-* (88 nodes) and *-lnd-gar-nodl-it (65 nodes).

			\begin{figure}[!bp]
					\centering
					\includegraphics[width=0.9\columnwidth]{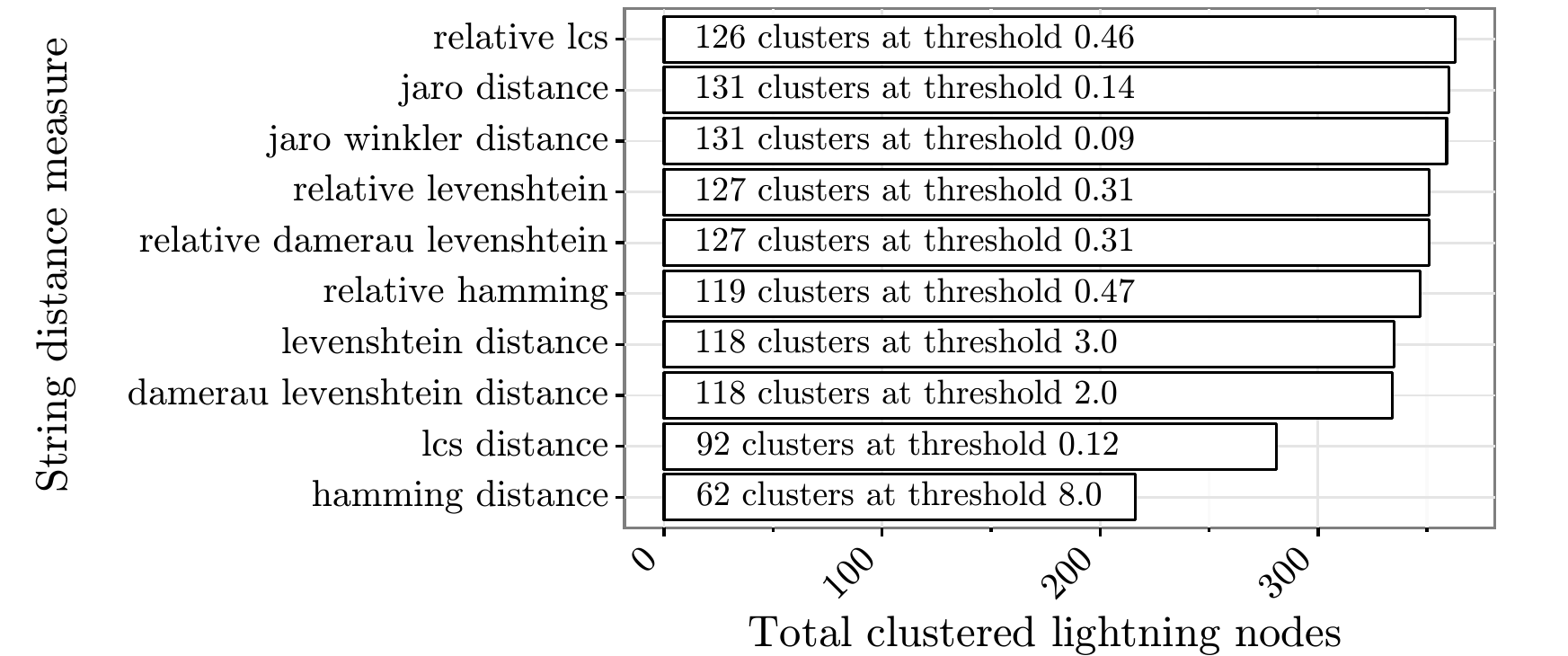}
					\caption{Comparison of string distance measures for alias clustering. The relative longest common substring (lcs) measure performs best. It grouped 363 \LN nodes into 126 clusters. The threshold of 0.46 implies that for two aliases to be clustered, their longest common substring needs to account for about half of the longer alias.}
					\label{fig:alias_clustering_results}
			\end{figure}

    \paragraph{Discussion}
        Alias/ASN and IP-based clustering can yield some false positives. For example, if two nodes have very similar aliases, and are coincidentally hosted on the same AS, they would be recognized as one entity. This could happen with \LN-specific hosting services or widespread services such as Amazon. Of the 301 identified clusters, 20 are hosted on Amazon servers, but are identified as distinct clusters due to their different naming schemes. Within the overall time frame of our dataset, 313 (2.9\%) different lightning nodes have at some point in time been hosted with Amazon.
        In general, however, filtering alias clusters to those running on the same AS should result in few false positives.
        For the entity LNBig.com, we had ground truth which we used to optimize the alias similarity threshold. By reaching out to one operator, we were able to validate one cluster of \LN nodes. For privacy reasons, we refrain from naming the operator.

    \paragraph{Countermeasures}
        While the use of aliases supports the usability of the system, the way some users choose them clearly hinders their privacy. For more privacy, aliases should be sufficiently different from one another. While the public announcement of IP addresses may be unavoidable for those nodes that wish to have incoming channels in the \LN, linkability across nodes of the same user can be mitigated if the clients for each node are hosted with different service providers (and thus ASNs and IP addresses).

\subsection{Evaluation of Different String Distance Measures\label{sec:evaluation-string-distance}}

	As illustrated in~\cref{fig:alias_clustering_results}, we compare the string measures lcs (longest common substring), Jaro, Jaro-Winkler, Levenshtein, Damerau-Levensthein and Hamming distance. For those distances where the result is not already between $0$ and $1$, we normalize the distance by dividing by the longer one of the two aliases to be compared.

	For example, a popular string edit distance, the Levensthein distance, measures the minimum number of single character edits that are needed to transform one string into another. Here an edit, refers to replacement, insertion or deletion.
	For a detailed overview on text similarity approches we refer the reader to~\cite{gomaa2013survey}.

	\begin{figure}[!bp]
			\centering
			\includegraphics[width=\textwidth]{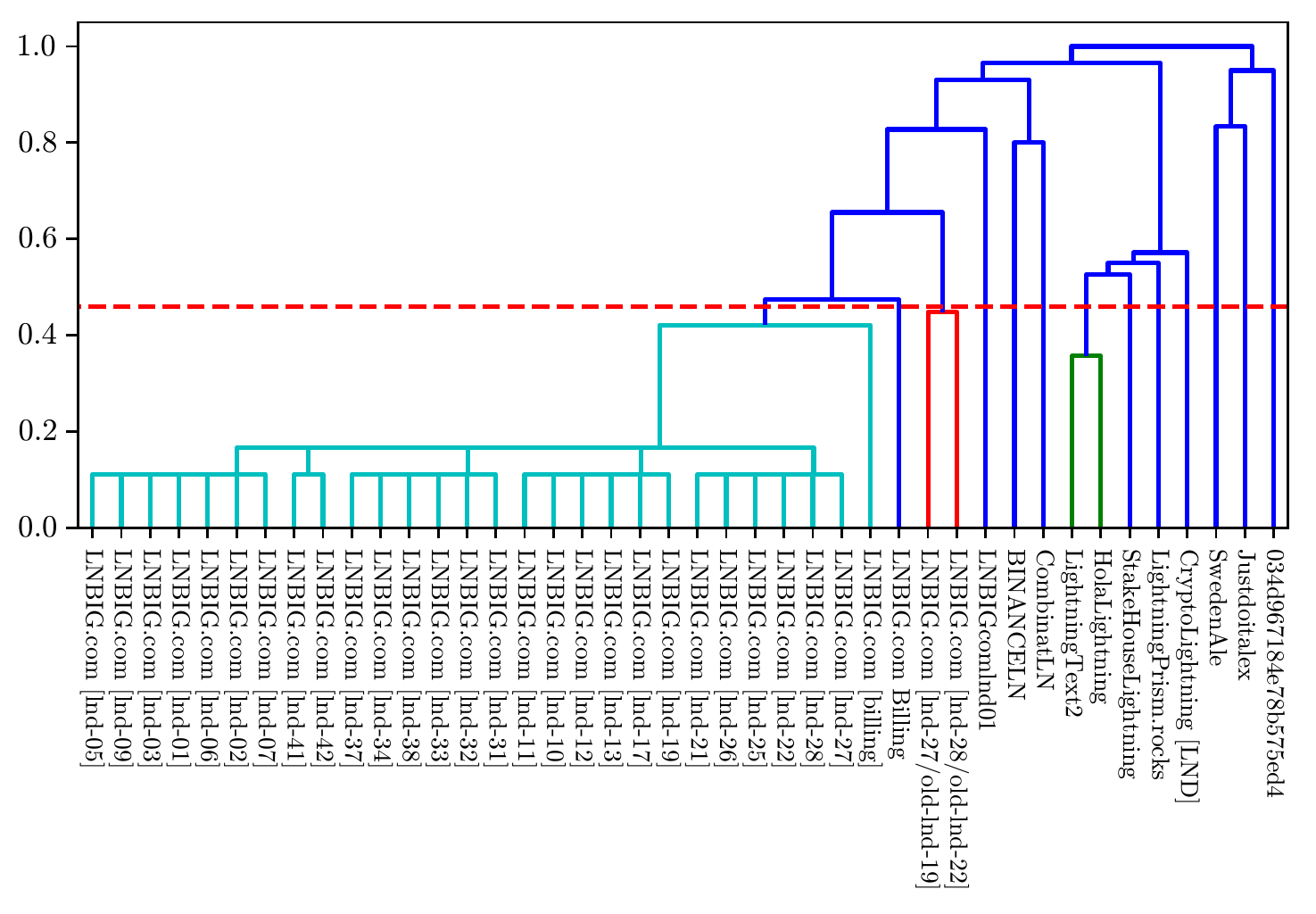}
			\caption{Example of a dendrogram illustrating alias similarity. Here, the relative lcs distance metric has been used along with the optimal threshold of 0.46 (red vertical dashed line). As a result, 3 clusters are found. However, only in the LNBig clusters, all nodes are hosted on the same AS. Node ids behind the 2 LNBig clusters overlap. Therefore, in this example, one LNBig.com cluster is the result.}
			\label{fig:alias_dendrogram_example}
	\end{figure}

	The results indicate that several normalized string distances exhibit similar performance, while the relative longest common substring yields the best performance. The optimal threshold of 0.46 can be interpreted as follows: if a common substring is identified between two aliases, it needs to account for about half of the length of the longer alias. A practical similarity comparison is illustrated in~\cref{fig:alias_dendrogram_example}. We have chosen a subset of aliases that contains all observed aliases of LNBig.com, multiple nodes containing the substring Lightning, and some randomly selected aliases. At the threshold, 3 clusters are identified. In two of them, all nodes are hosted on the same AS. So the initial result would be 2 identified clusters. As the cluster consisting of LNBIG.com [lnd-27/old-lnd-19] and LNBIG.com [lnd-28/old-lnd-22] are just additional aliases that have been seen over time, but actually belong to some of the same public keys of the other LNBig cluster, the two clusters are joined.

\section{Illustrative Example Linking Algorithm}
\label{sec:examplelinking}

\begin{figure}[tb]
			\centering
			\includegraphics[width=.5\columnwidth]{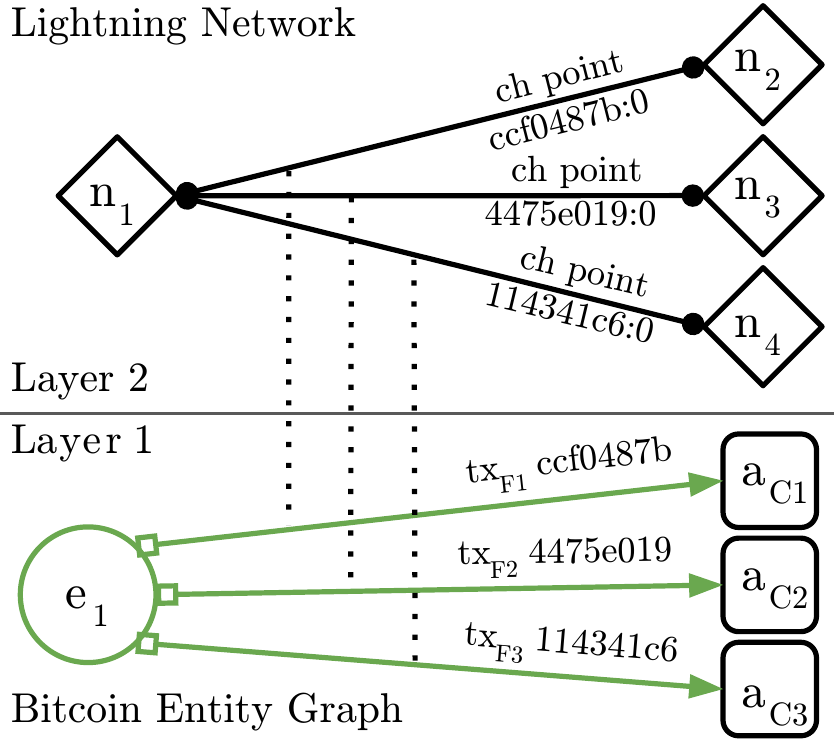}
			\caption{\textbf{\Algorithmlink 2: Entity reuse example.} At layer 1, the funding entity $e_1$ is reused to perform $N_C = 3$ \openchtxs. At layer 2, the corresponding channels are opened and there is one node, $n_1$, common to all the $N_C$ channels and it can be linked to the funding entity $e_1$.}\label{fig:entity_reuse_heuristic}
		\end{figure}

An illustrative example of the \algorithmlink described in~\cref{sec:heuristic2} is shown in~\cref{fig:entity_reuse_heuristic}, where \entity $\entityid_1$ funds $\numberofchannels := 3$ channels and a \node $\nodeid_1$ is common to all those channels. Then we can say that entity $e_1$ controls $n_1$.

\section{Details for Combining Heuristics}
\label{app:heuristics-contribution}

 In the best case (last entry in~\cref{tab:linking_clustering_results_diff}) we get to \percAddressesLinkedHeurTwoMinConfAll\% of linked addresses and \percNodesLinkedHeurTwoMinConfAllAliasIP\% of linked \LN nodes. 

	The reason why on-chain \algorithmscluster should improve the \algorithmslink is that they better represent a user's behavior, just like the co-spend heuristic does. If we think about the on-chain patterns that we introduced, they all group together entities that, based on their interaction with the \LN, are controlled by one single actor.

	\cref{tab:linking_clustering_results_diff} shows the percentage of addresses, entities and nodes that can be linked together when adding the \algorithmscluster in the linking process. Comparing these results with the ones from the basic implementation of the \algorithmslink, we notice that the first \algorithmlink improves only by few percentage points, while the second \algorithmlink improves roughly by a factor of 2. 
    However, not every \algorithmcluster contributes the same to the overall results. We discuss each of them next.

	\begin{table}[tbp]
		\centering
		\caption{Summary results }\label{tab:linking_clustering_results_diff}
		\adjustbox{max width=\columnwidth}{%
		\begin{tabular}{@{}lrrr@{}}
			\toprule
			Linking + Clustering                         & \multicolumn{1}{c}{\begin{tabular}[c]{@{}c@{}}\% addresses\\ linked\end{tabular}} & \multicolumn{1}{c}{\begin{tabular}[c]{@{}c@{}}\% entities\\ linked\end{tabular}} & \multicolumn{1}{c}{\begin{tabular}[c]{@{}c@{}}\% nodes\\ linked\end{tabular}} \\ \midrule
		    \rowcolor{Gray} \Algorithmlink 1 & \percAddressesLinkedHeurOneNone & \percEntitiesLinkedHeurOneNone & \percNodesLinkedHeurOneNone \\
			\Algorithmlink 1 + stars      & \percAddressesLinkedHeurOneStar & \percEntitiesLinkedHeurOneStar & \percNodesLinkedHeurOneStar \\
			\rowcolor{Gray} \Algorithmlink 1 + snakes     & \percAddressesLinkedHeurOneSnake & \percEntitiesLinkedHeurOneSnake & \percNodesLinkedHeurOneSnake \\
			\Algorithmlink 1 + collectors & \percAddressesLinkedHeurOneCollector & \percEntitiesLinkedHeurOneCollector & \percNodesLinkedHeurOneCollector \\
			\rowcolor{Gray} \Algorithmlink 1 + proxies & \percAddressesLinkedHeurOneProxy & \percEntitiesLinkedHeurOneProxy & \percNodesLinkedHeurOneProxy \\
			\Algorithmlink 1 + all on-chain & \percAddressesLinkedHeurOneAll & \percEntitiesLinkedHeurOneAll & \percNodesLinkedHeurOneAllAliasIP \\
			\rowcolor{Gray} \Algorithmlink 1 + all on/off-chain & \percAddressesLinkedHeurOneAll & \percEntitiesLinkedHeurOneAll & \percNodesLinkedHeurOneAll \\
			\Algorithmlink 2 & \percAddressesLinkedHeurTwoMinConfNone & \percEntitiesLinkedHeurTwoMinConfNone & \percNodesLinkedHeurTwoMinConfNone \\	
			\rowcolor{Gray} \Algorithmlink 2 + stars      & \percAddressesLinkedHeurTwoMinConfStar & \percEntitiesLinkedHeurTwoMinConfStar & \percNodesLinkedHeurTwoMinConfStar \\
			\Algorithmlink 2 + snakes    & \percAddressesLinkedHeurTwoMinConfSnake & \percEntitiesLinkedHeurTwoMinConfSnake & \percNodesLinkedHeurTwoMinConfSnake \\
			\rowcolor{Gray} \Algorithmlink 2 + collectors & \percAddressesLinkedHeurTwoMinConfCollector & \percEntitiesLinkedHeurTwoMinConfCollector & \percNodesLinkedHeurTwoMinConfCollector \\ 
			\Algorithmlink 2 + proxies & \percAddressesLinkedHeurTwoMinConfProxy & \percEntitiesLinkedHeurTwoMinConfProxy & \percNodesLinkedHeurTwoMinConfProxy \\ 
			\rowcolor{Gray} \Algorithmlink 2 + all on-chain & \percAddressesLinkedHeurTwoMinConfAll & \percEntitiesLinkedHeurTwoMinConfAll & \percNodesLinkedHeurTwoMinConfAll\\
			\textbf{\Algorithmlink 2 + all on/off-chain} & \percAddressesLinkedHeurTwoMinConfAll & \percEntitiesLinkedHeurTwoMinConfAll & \percNodesLinkedHeurTwoMinConfAllAliasIP\\
			\bottomrule
		\end{tabular}
		}
	\end{table}

	\paragraph{Star-pattern contribution} 
		The behavior that can be modeled when combining the star pattern and the \algorithmslink can be described with the following example. A user owns a wallet and additionally controls one \LN node $n$ which runs its own \LN wallet. Anytime the \LN wallet needs to be replenished, it generates a different address $a_i$ (corresponding to an entity $e_i$ of size 1) and the wallet sends coins to it. After this, $e_i$ can be used to open a new channel from the node $n$. At this point, the node $n$ can be linked to the star that is formed by the set of entities $\set{e_i}$. 

		Unfortunately, this pattern, as reported in~\cref{tab:clustering_results_onchain}, occurs less often than the others, a possible reason why it has no contribution for the \algorithmlink 1 and an impact of less than a percentage point in \algorithmlink 2.

	\paragraph{Snake-pattern contribution}
		As already described in~\cref{sec:clustering_on_chain} the snake pattern follows the concept of reusing the change address to fund a new channel. Due to the frequent creation of a change in \Bitcoin, this pattern occurs much more often than the star pattern and the proxy pattern (two and one order of magnitudes more respectively). This also the reason why its contribution to the linking is the most significant one for \algorithmlink 2. Unfortunately, it is not so effective with the \algorithmlink 1, probably because the coin-reuse heuristic is a stricter version of the entity-reuse heuristic. 

	\paragraph{Proxy-pattern contribution}
		The proxy pattern models the behavior of an \LN user that decides to merge the coins from different \closechtxs into one single entity to avoid keeping track of funds, possibly on different wallets.
		This pattern seems to have a stable contribution (around 3\% for linked nodes) for both \algorithmslink when applied without the other patterns.

	\paragraph{Collector-pattern contribution}
		Similar to the proxy pattern, this behavior merges the coins from different \closechtxs into one single entity, with the difference that this last one is not directly involved in the \LN settlements. This pattern appears to be less common and powerful compared to the proxy pattern.

	\paragraph{Off-chain node clustering contribution}
        Assume there is a cluster of nodes obtained with the heuristic presented in~\cref{sec:clustering_off_chain} and one of these nodes has been linked to one entity. At this point, since the nodes in the cluster are supposed to be controlled by the same \LN user, we can indirectly link all the other nodes in the cluster to the entity. We refer to these nodes as \emph{indirectly-linked nodes}. Even though we enforced strict conditions in the \algorithmcluster based on alias/IP information, we are aware of the fact that this type of linking may be considered weaker as it relies on one additional assumption (nodes in an alias-based cluster are correctly attributed to one actor).
		In total, for the \algorithmlink 1 we find \nNodesLinkedAliasIPClusterHeurOne indirectly-linked nodes, which corresponds to an additional \percNodesLinkedAliasIPClusterHeurOne\% of nodes linked, while for the \algorithmlink 2 we find \nNodesLinkedAliasIPClusterHeurTwo indirectly-linked nodes which correspond to an additional \percNodesLinkedAliasIPClusterHeurTwo\% of nodes linked.

\section{Security and Privacy Implications\label{app:secanalysis}}

In this section, we evaluate the security and privacy implications of our clustering and linking algorithms in the  security and privacy impact on the \LN.

        \begin{table}[t]
            \vspace{-0.3cm}
                \centering
                \caption{\LN users controlling most capacity}\label{tab:distribution-capacity}
                \adjustbox{max width=\columnwidth}{%
                \begin{tabular}{@{}lrr@{}}
                    \toprule
                    User             & Node count & Share of total capacity contributed \\ \midrule
                    \rowcolor{Gray} LNBig.com *        & 26         & 24.07\% \\
                    bfx-lnd*                           & 2          & 4.20\% \\
                    \rowcolor{Gray} BitRefill.com, ... & 3          & 2.37\%  \\
                    CoinGate                           & 2          & 1.98\% \\
                    \rowcolor{Gray} Breez              & 3          & 0.52\% \\
                    \bottomrule
                \end{tabular}
                }
        \end{table}

    \subsection{Wealth Distribution and Impact of Griefing Attacks in the \LN\label{sec:wealth}}
        In this section, we first evaluate how wealth is distributed in the \LN, that is, how much capacity is controlled by each of the users found during our analysis. For that, we take a recent snapshot of the \LN from 2020-09-09 and extract the capacity controlled by each user. If a channel has been created by a user that has been linked to a node, we can attribute the full capacity of the channel to that node.
        For all other channels, we assume that each user controls half of the capacity. Under these assumptions, we observe that the overall capacity of the \LN is distributed as shown in~\cref{tab:distribution-capacity}. In particular, a single user controls over 24\% of the overall capacity in the \LN and as few as 237 nodes (3.2\%) control over 80\% of the capacity. This result refines the previous study  in~\cite{Lin:2020a} where they find that 80\% of the capacity is controlled by 10\% of the nodes.

        This result shows that few \LN users are in a privileged situation that they can potentially use to selectively prevent other \LN nodes from transacting in the network, for instance, launching a \emph{griefing attack}~\cite{griefing-attack} against the victim nodes. In the griefing attack, the attacker finds a path of the form $\nodeid_1 \rightarrow \nodeid_2 \rightarrow \ldots \rightarrow \nodeid_k$ where $\nodeid_1$ and $\nodeid_k$ belong to the attacker. Using that path, the attacker routes a payment from $\nodeid_1$ to $\nodeid_k$, thereby allocating funds at each channel to support the payment transfer. However, this payment is never accepted by $\nodeid_k$, forcing the intermediary channels to wait to release the funds locked for the payment until a certain timeout expires. In the current \LN implementation, this timeout is in the order of several days.

\begin{figure}[!tbp]
          \centering
          \begin{minipage}[b]{0.49\textwidth}
            \includegraphics[width=\textwidth]{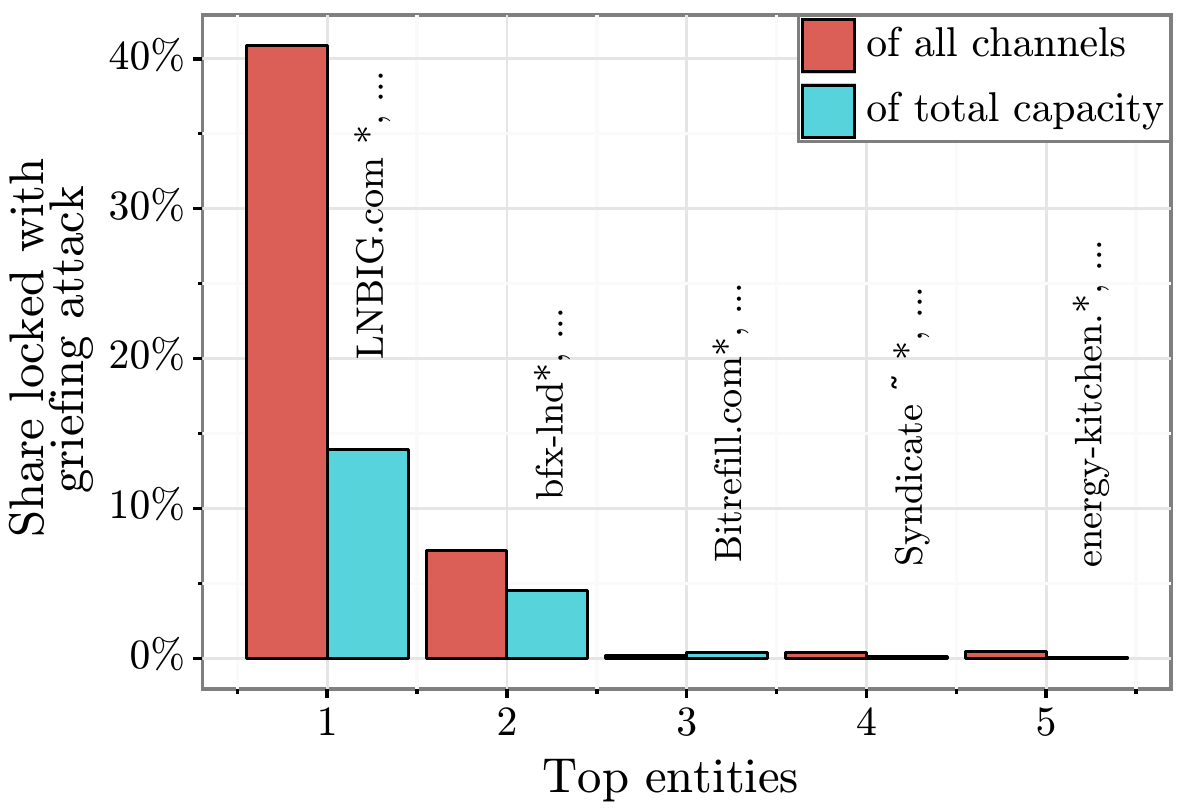}
            \caption{Fraction of all \LN channels and capacity vulnerable to a griefing attack launched by most effective entity.}
            \label{fig:grieffing-attack}
          \end{minipage}
          \hfill
          \begin{minipage}[b]{0.49\textwidth}
            \includegraphics[width=\textwidth]{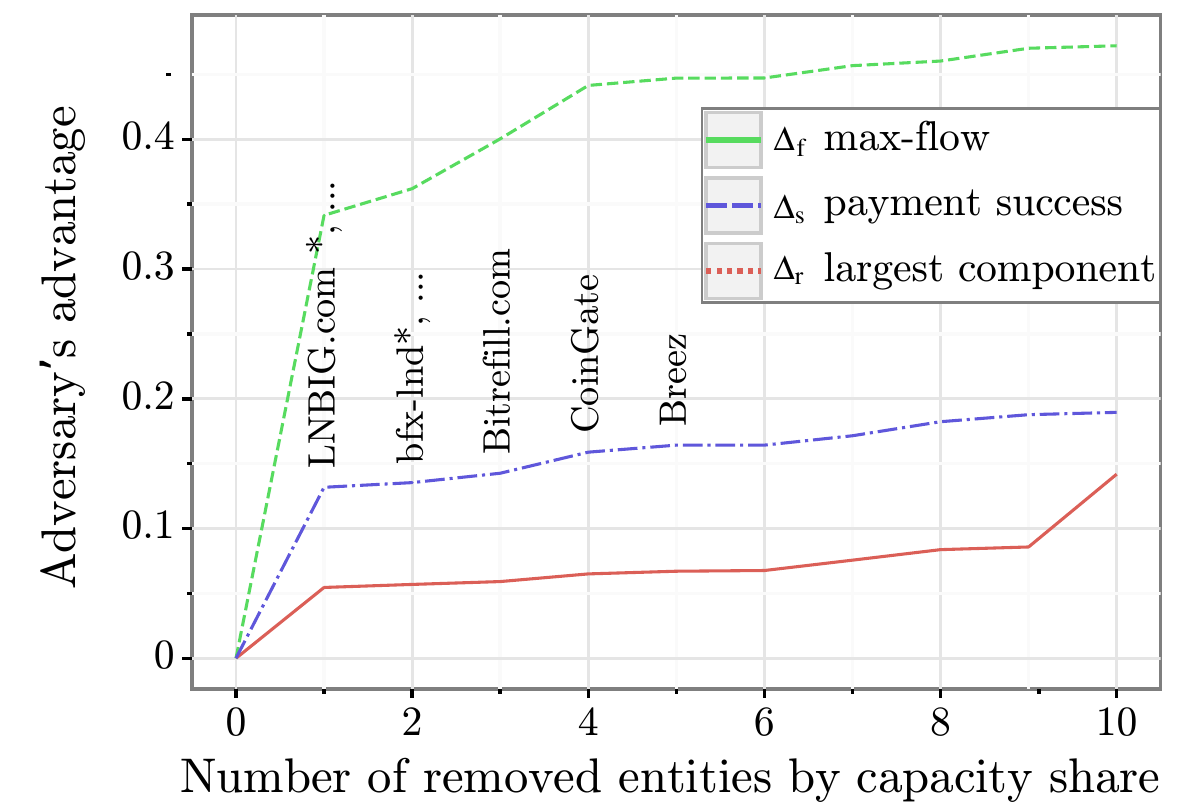}
            \caption{Adversary's advantage in impeding max-flow, payment success and largest component size --- given an entity DoS.}
            \label{fig:dos-attack}
          \end{minipage}
        \end{figure}

        For this attack to be effective, the attacker needs to perform and lock a payment for an amount corresponding with the capacity available at the channel of the victim. However, as shown in~\cref{tab:distribution-capacity}, the uneven distribution of wealth in the \LN makes this a small investment if the attacker is one of the users with high capacity. In fact, we evaluated the possible damage that each user in the \LN can infringe by launching this griefing attack with the results shown in~\cref{fig:grieffing-attack}. As expected from the wealth distribution, the user with the highest capacity is the one that can infringe the most devastating attack, being able to render useless for a period of time over $40\%$ of the channels in the \LN, which amount for about $14\%$ of the total capacity.

        We remark that although griefing attacks have a cost for the adversary (i.e., the adversary needs to lock some of its own channels), the adversary can still benefit from griefing other nodes, as studied in the literature. For instance, P\'erez-Sol\`a et al.~\cite{PerezSola:2019} show that the adversary can launch a griefing attack to block \LN middle nodes in multi-path payments. Mizrahi and Zohar~\cite{Mizrahi:2020} as well as Rohrer et al.~\cite{Rohrer:2019aa} show how similar attacks can be used to block as many high liquidity channels as possible, disconnect channels from the \LN and isolate individual nodes from the \LN.
        If the adversary is successful, the attack gives the adversary a  dominant  position  in  the  \LN,  which  can  be  later  exploited  either  for exploiting  privacy (e.g., off-chain payment data gathering)  or  for economic rewards (e.g., increasing  the  benefits  in term of fees reducing the number of competing \LN  gateway nodes).

    \subsection{Vulnerability to DoS attacks in the \LN\label{sec:dos-attacks}}
        The growing monetary value of the \LN and the existence of competitor business within the network as well as from other available payment networks open the door for DoS attacks. In fact, there have already been DoS attacks against the \LN reported. For instance, in March 2018, it was reportedly hit by a distributed DoS attack that took $20\%$ of the nodes offline\footnote{\url{https://trustnodes.com/2018/03/21/lightning-network-ddos-sends-20-nodes}}. In this state of affairs, we study here the effect of DoS attacks targeted at the \LN users found in this work.

        Based on the \LN snapshot we iteratively remove the nodes and channels corresponding to a given user, starting with the users that control the most capacity. We then compare the resulting graph with the original one to evaluate the \emph{adversary's advantage} (i.e., attack's success) attributed to a DoS targeted to such user. Following~\cite{Rohrer:2019aa}, we characterize the notion of adversary's advantage as $\Delta_m := \left| 1 - \frac{m'}{m}\right|$ where $m$ is the a priori measurement and $m'$ is the a posterior one. The higher $\Delta_m$ becomes, the higher the success of the attack according to the metric $m$. We consider the three metrics as defined in~\cite{Rohrer:2019aa}:  (i) $\Delta_r$ defined as the number of nodes within the biggest component in the graph, representing thereby the effect on the number of reachable nodes; (ii) $\Delta_f$ defined as the average maximum flow between every two nodes in the graph, representing thereby the effect of the attack on liquidity; and (iii) $\Delta_s$ defined as the payment success ratio, representing thereby the effect of the attack on the payments.
        Following their approach for estimation, we perform a uniform random sampling of 1000 pairs of  nodes to compute $\Delta_f$ and $\Delta_s$.

        We obtain the results shown in~\cref{fig:dos-attack}. We observe that a possibly low resource adversary that carry out a DoS attack targeted to a single \LN user (LNBig.com, 26 nodes in total) already gets an advantage that is only slightly improved when targeting more users. By attacking this entity, the max-flow of the \LN can be reduced by one third, and payment success be reduced by 12\%. As each user is hosted on a single autonomous system, it could be sufficient to attack a single hosting provider. In this regard, our results differ from those in~\cite{Rohrer:2019aa}. Multiple high degree nodes are likely using several hosting providers, increasing the attack's cost. Second, even with a lower budget requirement, our DoS attack targeted at users yields a similar adversary's advantage as in~\cite{Rohrer:2019aa} for all the metrics when only considering one user with 26 nodes.

    \subsection{\LN Users on Payment Paths\label{sec:across-paths}}
        In this section, we study to what degree the security and privacy of individual payments between any two nodes in the \LN are affected by our clustered users. In the \LN, a payment between two nodes is typically routed through the cheapest path between them, where the cost associated to the path is calculated by the sum of fees charged by each intermediary node. An intermediary node charges a fee composed of
        a rate fee proportional to the payment amount, and a base fee that is independent. We computed the cheapest paths between all node pairs for a varying payment amount.
        This allows us to study the value privacy property, that is visualized in~\cref{fig:value-privacy-and-rel-anonymity}.

        \paragraph{Value privacy}
            The payment value is observed by every single intermediary in the path. Thus, according to our results, a reduced number of users know how many coins are being transferred in the \LN, giving them undue advantage over competitors (e.g., to set the fees or target products to users accordingly).
            Being an intermediary also has a second implication, from a security point of view: a payment between any two nodes can be aborted by a single intermediary node that simply drops it. A single user can thus stop almost $40\%$ of the payments in the \LN, and this fraction grows to $60\%$ if the top 5 users were to collude.
            Given the decentralized payment protocol used currently in the \LN, it is not possible for the sender to pinpoint which intermediary node has stopped the payment. Therefore the sender needs to blindly guess what node is the malicious one and possibly pay higher fees to circumvent it.

    \subsection{\LN Users With Multiple Nodes on Payment Paths\label{sec:within-path}}
        From the results in the previous section, we observe that a few users are frequently intermediary nodes for many paths used for payments in the \LN. In this section, we are interested in studying whether a single user has more than one node as an intermediary in a single path. This setting has further security and privacy implications in practice.

        \paragraph{Relationship anonymity}
            Assume a path where a user has two nodes, one of which is the immediate successor of the sender and the other is the immediate predecessor of the receiver. In such a setting, the fact that information uniquely identifying a payment is sent across the path (e.g., a hash value used to cryptographically secure the payment), allows the user to learn the sender and receiver for such payment, even when other simultaneous payments may be using part of the path. This privacy attack breaks the notion of \emph{relationship anonymity} as described in~\cite{Malavolta:2017a}.

\begin{figure}[!tbp]
              \centering
              \begin{minipage}[b]{0.49\textwidth}
                \includegraphics[width=\columnwidth]{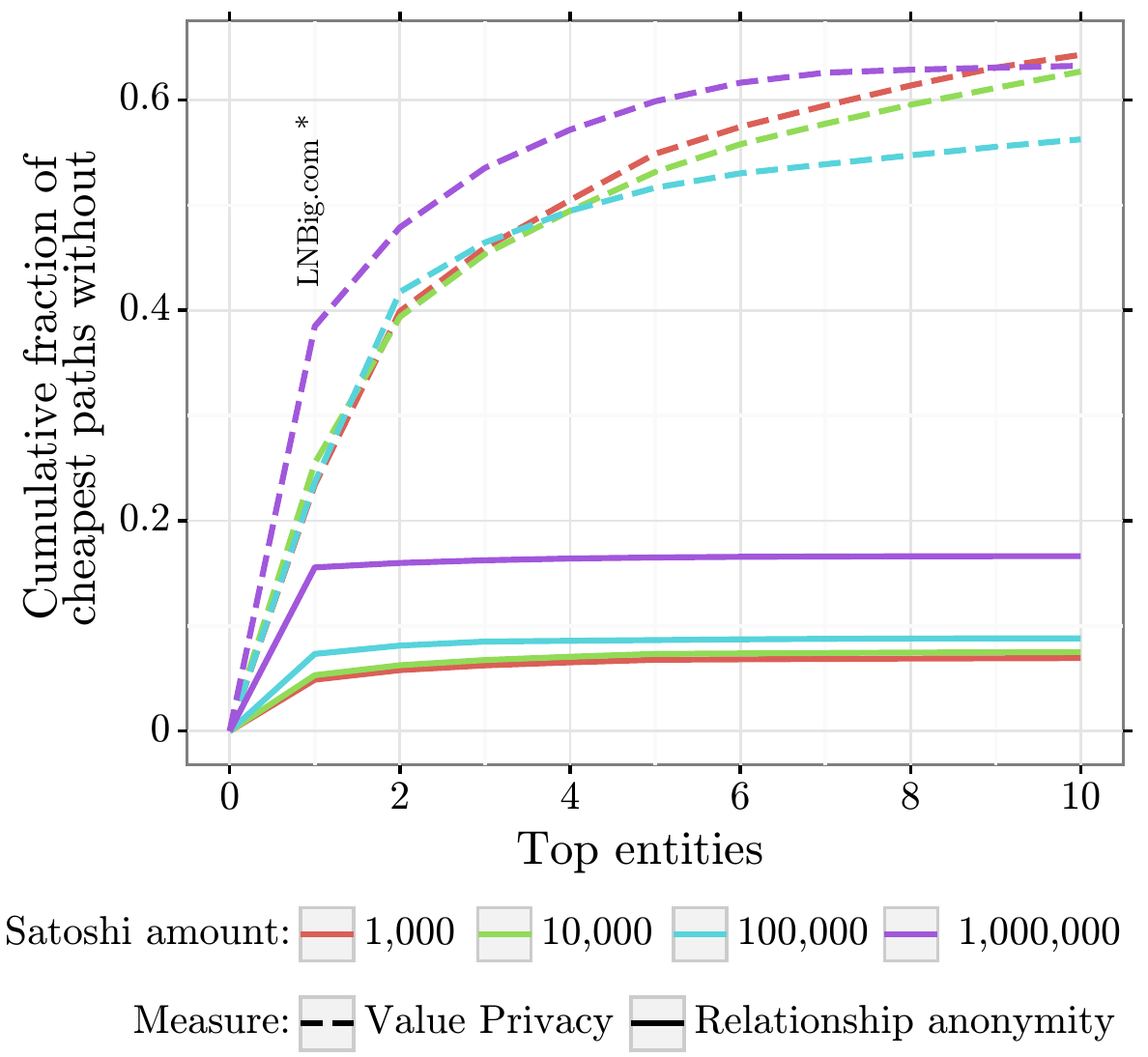}
                \caption{Fraction of cheapest paths without value privacy and relationship anonymity by different amounts to be transferred.}
                \label{fig:value-privacy-and-rel-anonymity}
              \end{minipage}
              \hfill
              \begin{minipage}[b]{0.49\textwidth}
                \includegraphics[width=1\columnwidth]{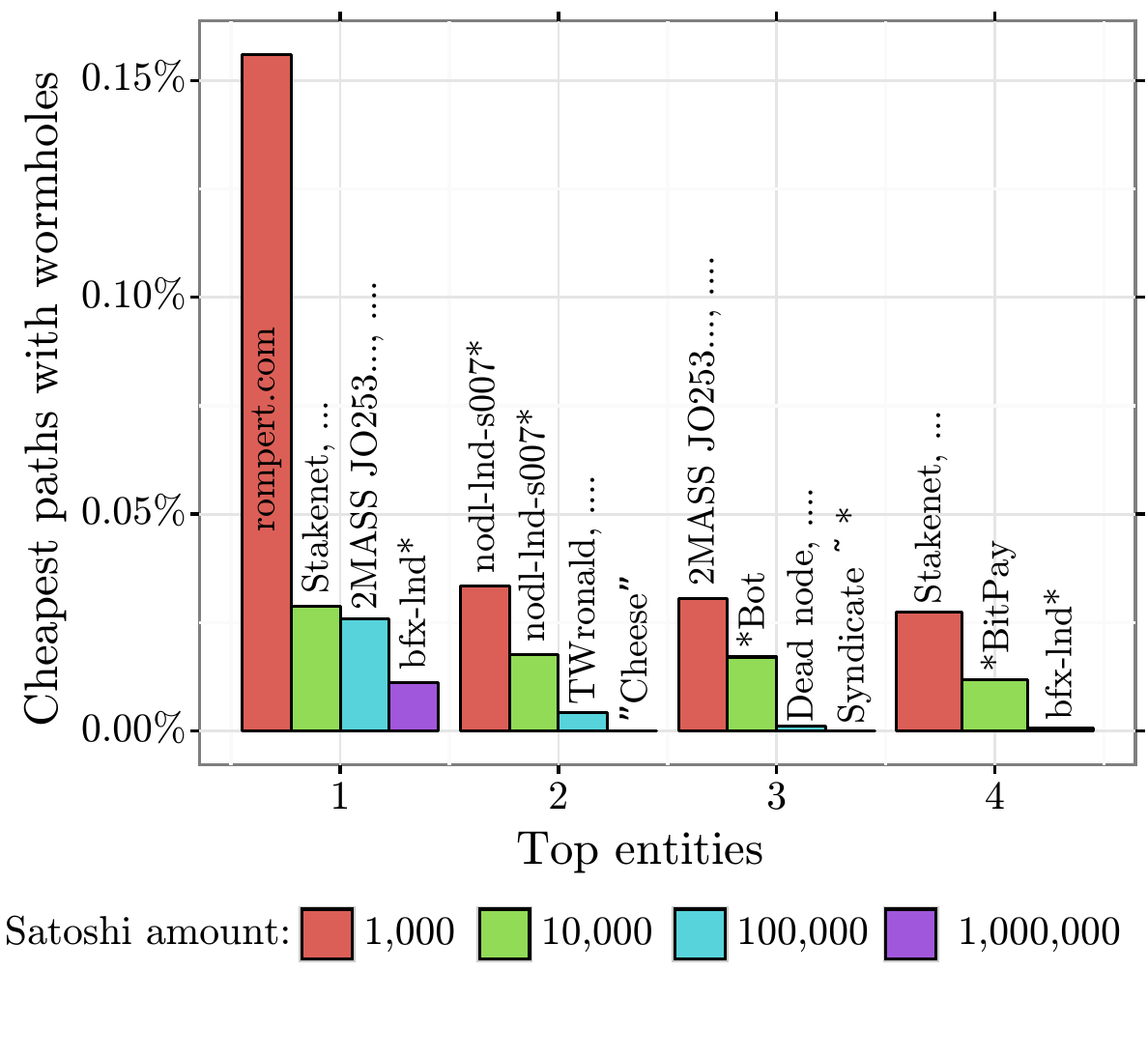}
                \caption{Cheapest paths prone to the wormhole attack. In theory, the rompert.com entity could exploit the most wormholes.}
                \label{fig:wormhole}
              \end{minipage}
            \end{figure}

            We evaluated the presence of such a threat in the \LN with the results shown in~\cref{fig:value-privacy-and-rel-anonymity}. We observe that  there are between $5\%$ and $16\%$ of the paths prone to this privacy issue even when as little as one user behaves adversarial. The reason why relationship anonymity is much more vulnerable for higher payment amounts is straightforward: Only a few channels have sufficient capacity, and several of them are operated by the same user (i.e. LNBig.com), forcing more payments to go through them.

        \paragraph{Wormhole attack}
            Assume now a path where a user has two intermediary nodes at any position in the path with the condition that there are other honest nodes between them. The latter are at risk of becoming a victim of the \emph{wormhole attack} as described in~\cite{Malavolta:2019a}. They are tricked into locking capacity at their channels to facilitate the payment but never contacted again to release those funds so that channels get locked for a certain timeout period established as system parameter which is on the order of days in the current implementation. While similar in spirit to the griefing attack, the wormhole attack differs in two main points: (i) the attacker user does not need to be the sender and receiver of the payment; and (ii) the attacker user can successfully settle the payment at the channels in the path other than those being attacked (i.e., channels between two nodes of the attacker), so that the attacker also gets the fees for providing an apparently successful payment at the eyes of the sender and the receiver.

            As shown in~\cref{fig:wormhole}, surprisingly the user with the highest impact in this attack is not LNBIG.com as in the previous attacks. In this case, the user associated to rompert.com can perform the wormhole attack for about $0.15\%$ of all cheapest paths in the \LN. While this number is much lower than in previous attacks, the effect of this attack actively disrupts users in the path (i.e., their coins get locked), different to privacy-based attacks where the payment finishes successfully and the privacy breach is computed locally and passively at the attacker node.

	These results call for the inclusion into the current \LN of countermeasures recently proposed in the literature. For instance, Egger et al.~\cite{Egger:2019} and Malavolta et al.~\cite{Malavolta:2019a} have proposed alternative payment schemes for the \LN that prevent the worhmole attack. Moreover, the payment schemes proposed by Malavolta et al.~\cite{Malavolta:2017a,Malavolta:2019a} and Yu et al.~\cite{Yu:2019} provably prevent the relationship anonymity attack that are otherwise currently feasible in the \LN.

    \subsection{The Good and the Bad for Routing in the \LN\label{sec:routing}}
        The possibility of deanonymization, which opens up with the cluster and linking algorithms proposed in this work, has the following implications arising from the security and privacy issues in the routing of payments discussed so far.

        Honest users can use the knowledge about users to search for payment paths that avoid them. However, this may not always be possible, especially for users who control a node with only a few channels. In addition, alternative paths circumventing these users may be more expensive, which represents a trade-off between security/privacy and transaction fees.

        On the other hand, the fact that honest users can learn about users and avoid them may have a negative impact on the business model of these users. The business incentive for the \LN nodes is to offer many channels and to set their fees so that as many payments as possible are routed through them. More payments are also associated with higher revenue potential. From this point of view, the deanonymization techniques presented in this work are not beneficial for routing users.

\end{document}